\documentclass[a4paper,fleqn,usenatbib]{mnras}

\usepackage{graphicx}	
\usepackage{amsmath}	
\usepackage{amssymb}	
\usepackage{natbib}

\title[Compacts Quench Faster than Normal-sized Galaxies]{Compact Galaxies at Intermediate Redshifts Quench Faster than Normal-sized Galaxies}
\author[Nogueira-Cavalcante et al.]{J.~P.~Nogueira-Cavalcante$^{1,2}$\thanks{E-mail: jpncavalcante@on.br}, T. S. Gon\c{c}alves$^{1}$, K. Men\'endez-Delmestre$^{1}$,
\newauthor I. G. de la Rosa$^{3,4,5}$ and A. Charbonnier$^{1}$\\
$^{1}$Observat\'orio do Valongo, Universidade Federal do Rio de Janeiro, Ladeira Pedro Ant\^onio 43, Sa\'ude 20080-090 Rio de Janeiro, RJ, Brazil\\
$^{2}$Observat\'orio Nacional, Rua Gal. Jos\'e Cristino 77, S\~ao Crist\'ov\~ao 20921-400 Rio de Janeiro, RJ, Brazil\\
$^3$ Instituto de Astrof\'\i sica de Canarias, C/ V\'\i a L\'actea s/n, La Laguna, E-38200 La Laguna, Tenerife, Spain\\ 
$^4$ Departamento de Astrof\'\i sica, Universidad de La Laguna, E-38205 La Laguna, Tenerife, Spain \\
$^5$ National Astronomical Observatory of Japan, Osawa, Mitaka, Tokyo 181-8588, Japan \\
}
\date{Accepted XXX. Received YYY; in original form ZZZ}

\pubyear{2019}

\begin{document}
\hypersetup{pageanchor=true}
\label{firstpage}
\pagerange{\pageref{firstpage}--\pageref{lastpage}}
\maketitle

\begin{abstract}

Massive quiescent compact galaxies have been discovered at high redshifts, associated with rapid compaction and cessation of star formation (SF). In this work we set out to quantify the time-scales in which SF is quenched in compact galaxies at intermediate redshifts. For this, we select a sample of green valley galaxies within the COSMOS field in the midst of quenching their SF at $0.5<z<1.0$ that exhibit varying degrees of compactness. Based on the H$\delta$ absorption line and the 4000 \AA \ break of coadded zCOSMOS spectra for sub-samples of normal-sized and compact galaxies we determine quenching time-scales as a function of compactness. We find that the SF quenching time-scales in green valley compact galaxies are much shorter than in normal-sized ones. In an effort to understand this trend, we use the Illustris simulation to trace the evolution of the SF history, the growth rate of the central super massive black hole (SMBH) and the AGN-feedback in compact and normal-sized galaxies. We find that the difference in SF quenching time-scales is due to the mode of the AGN-feedback. In the compact galaxies the kinematic-mode is dominant, being highly efficient at quenching the SF by depleting the internal gas. For normal-sized galaxies, the prevailing thermal-mode injects energy into the circumgalactic gas, preventing cold gas accretion and quenching SF via the slower strangulation mechanism. These results are consistent with the violent disk instability and gas-rich mergers scenarios, followed by strong AGN and stellar feedback. Although this kind of event is most expected to occur at $z=2-3$, we find evidences that the formation of compact quiescent galaxies can occur at $z<1$.

\end{abstract}




\section{Introduction}

The colour-magnitude diagram (CMD) of galaxies reveals a distinct peak in number density towards redder colours and a second one towards the bluer end. Recent studies \citep[e.g., ][]{Kauffmann2003} have pointed out the differences in physical properties between galaxies in the two peaks, revealing that blue galaxies (the so-called \textit{blue cloud} galaxies) typically have younger stellar populations than those in the red peak (\textit{red sequence} galaxies). With a marked extension towards higher stellar masses, these red sequence galaxies also dominate the high-mass end.

When the Galaxy Evolution Explorer \citep[GALEX; ][]{Martin2005} was launched, the near-UV band, sensitive to the peak of young ($< 100$ Myrs) stars, became key for the study of star-forming systems. Taking advantage of this, \citet{Wyder2007} demonstrated that a CMD based on NUV$-r$ colour $-$ combining GALEX NUV and Sloan Digital Sky Survey (SDSS) $r$ photometries $-$ was significantly more effective at separating the blue cloud and red sequence populations, revealing more clearly the region between these peaks. This region has been denominated as the \textit{green valley} \citep[e.g., ][]{Martin2007, Wyder2007, Goncalves2012, Salim2014, Schawinski2014, Smethurst2015, Nogueira-Cavalcante2018}. \citet{Pan2013} have shown that stellar populations in green valley galaxies are older than those in blue cloud galaxies and younger than those in red sequence galaxies, supporting an evolutionary scenario on the galaxy CMD, in which blue galaxies evolve into red ones, transitioning through the green valley. Therefore the green valley could be considered as a transitional phase, where star-forming galaxies are becoming quiescent systems. 

New episodes of star formation may be triggered in passive galaxies. For instance, \citet{Trayford2016} studied the $u-r$ colour evolution of simulated galaxies from EAGLE simulations and they found that, at $z<2$, up to $17\%$ of green valley simulated galaxies are red galaxies undergoing new star formation episodes. For these \textquotedblleft rejuvenated\textquotedblright \ galaxies \citet{Martin2017} and \citet{Darvish2018} argue that these systems could be examples of galaxy \textit{cannibalism} (accretion of smaller gas-rich galaxies by more massive galaxies). Such events are suggested as the origin of giant galaxies which reside at the centre of clusters \citep[e.g., ][]{Ostriker1975, White1976, Hausman1978, Malumuth1984}. However, for the bulk of the galaxy population, we can consider that galaxies generally evolve from star-forming systems to passive ones. 
 
The scarcity of galaxies in the green valley region suggested early-on that the transition from blue to red happens relatively fast. \citet{Goncalves2012} estimated that the mass flux through the green valley (i.e., the amount of stellar mass passing through the green valley per year and per unit of Mpc$^3$) is $\sim$0.16 M$_{\odot}$ yr$^{-1}$ Mpc$^{-3}$ at $z\sim0.8$. This is 5 times greater than the mass flux measured at $z\sim0.1$ \citep[0.033 M$_{\odot}$ yr$^{-1}$ Mpc$^{-3}$, ][]{Martin2007}. These results have demonstrated that galaxy transition on the CMD is faster at higher redshifts, indicating that the mechanisms capable of halting star-formation in galaxies are more efficient at earlier times. 

At a given mass, red sequence galaxies are typically more centrally-concentrated when compared to blue cloud galaxies \citep[e.g., ][]{Cheung2012, Fang2013, Woo2015, Williams2017}. This suggests that independently of which galaxy-wide agents (e.g., bar/spiral structures, AGN, merger, galaxy environment) transforms star-forming galaxies as they transition through the green valley, they must cause compaction in host galaxies. In other words, \textit{compaction is a necessary condition to galaxies becoming quenched} \citep[e.g, ][]{Fang2013}.

Many works have provided insights as to the mechanisms that may be at work in driving galaxies through the green valley and how they impact host galaxy properties. \citet{Nogueira-Cavalcante2018} found that at $0.5<z<1.0$ early-type galaxies have quenching time-scales twice as short as that of late types ($\sim150$ Myr and $\sim300$ Myr respectively). This demonstrates that quenching is a function of galaxy morphology, in line with hydrodynamical simulations that have also pointed to the dependency of quenching on morphology \citep{Martig2009,Martig2013}. At lower redshifts ($0.02<z<0.05$) \citet{Schawinski2014} and \citet{Smethurst2015} have also demonstrated that early-type galaxies quench star formation faster  than late-type galaxies ($\sim250$ Myr and $1-2$ Gyr, respectively). Although the methodologies in these works are different, the shorter quenching time-scales at higher redshifts by \citet{Nogueira-Cavalcante2018} point to the downsizing scenario, with more massive galaxies at earlier epochs quenching their star formation faster. 

In simulations, feedback as a quenching mechanism has also been shown to be critical to avoid overpredicting the galaxy abundance at low and high ends of the mass function: stellar feedback \citep[through winds from supernovae and massive stars, ][]{Menci2005, Hopkins2012, Lagos2013} and active galactic nuclei \citep[AGN, ][]{DiMatteo2005, Nandra2007, Cattaneo2009, Schawinski2009, Fabian2012, Dubois2013, Smethurst2016, Martin2017} are able to suppress star formation in low-mass and massive galaxies, respectively. Mergers can also play an important role in quenching star formation, specially at high redshifts \citep{Peng2010}, when galaxy interactions become more common \citep[e.g., ][]{Conselice2003, Lotz2011}. This mechanism is also associated to the formation of elliptical galaxies \citep[e.g., ][]{Toomre1977, Springel2005a}, although numerical simulations have also suggested that the disk component can remain after a major merger event \citep{Springel2005e, Querejeta2015, Pontzen2017, Sparre2017}. At lower redshifts, when mergers are less common, secular evolution $-$ associated with a galaxy's interaction with its immediate environment (e.g., satellites, prolonged gas infall) and gas-mixing via internal structures (stellar bars, spiral arms and oval distortions) $-$ may become more important in transitioning galaxies from star-forming to passive systems \citep{Kormendy2004, Masters2010, Masters2011, Cheung2013}. In fact, the most important secular agent is the stellar bar \citep{Kormendy2004}, being present in $\sim 67 \%$ of local disk galaxies \citep[e.g., ][]{Menendez-Delmestre2007}. 

Considering that compaction appears to be intimately linked with the transition from blue to red galaxies in the CMD, we expect a significant population of compact passive galaxies as the red sequence grows. The identification of a population of very compact massive galaxies at $z=2-3$ has shed light on this matter. These objects, called \textit{red nuggets}, are $3-5$ more compact than their counterparts in the local universe with the same mass \citep{Cimatti2004, Daddi2005, Trujillo2006, Damjanov2009, Damjanov2011, Whitaker2012, Cassata2013, Huertas-Company2016, Wel2014}. The abundance of red nuggets reaches a peak around $z\sim1.5$ and decreases at lower redshifts \citep{Cassata2013, Wel2014, Dokkum2015, Charbonnier2017}, suggesting that such objects either have been increasing their sizes and/or become the bulges of local galaxies \citep{Graham2015, delaRosa2016, Barro2017}. 

Although the redshift distribution of blue and red nuggets suggests that these galaxies transit through the green valley at $z>1$, in this work we identify a significant number of green valley compact galaxies (\textit{green nuggets}) at $0.5<z<1.0$. We aim to measure quenching time-scales of this population of green valley compact galaxies and compare them with those of normal-sized counterparts, with the objective of understanding whether compactness is correlated with faster quenching time-scales. We interpret our results applying this same analysis in simulated galaxies from the Illustris simulation. In the context of the present study, simulations serve a double purpose. On one hand, simulations reinforce observational findings by reproducing the observational results. On the other hand, simulations can be used to interpret the observational findings via their unique ability to track the individual galaxy evolution. 

We organise this paper as follows. In Section \ref{methodology} we present the methodology used to measure the star formation quenching time-scales based on galaxy spectra and colours. In Sections \ref{galaxy_sample} and \ref{illustris_simulation_description} we describe the galaxy data used in this analysis, the simulations and the criteria to determine the compactness of galaxies. In Section \ref{results} we present our results and in Section \ref{conclusions_and_discussion} we conclude and discuss our findings. Finally, in Section \ref{summary} we provide a summary of this work. Through this paper we use standard cosmology ($\Omega_\text{M}=0.3$, $\Omega_{\Lambda}=0.7$ and $h=0.7$) and rely on the AB magnitude system. 

\section{Measuring quenching time-scales}\label{methodology}

In this work we use the same methodology to measure quenching time-scales in the local universe as introduced by \citet{Martin2007}, and used at higher redshifts by \citet{Goncalves2012} and \citet{Nogueira-Cavalcante2018}. In the following paragraphs we briefly summarise the technique, but further details and caveats can be found in these works.

To quantify the quenching time-scales of the galaxies in our sample we assume that the star formation histories (SFHs) of green valley galaxies are described by a period of constant star formation rate (SFR), followed by a exponential decay:
\begin{equation}
  \text{SFR}(t)=\text{SFR}(t=0), \,\,\,\,\ t < t_c ;
  \label{star_formation_history_1}
 \end{equation}
 \begin{equation}
  \text{SFR}(t)=\text{SFR}(t=0)e^{-\gamma t}, \,\,\,\,\ t >  t_c \,\, ,
  \label{star_formation_history_2}
 \end{equation}
where $t_c$ is a characteristic time, in units of Gyr, and the $\gamma$ index, in units of Gyr$^{-1}$, determines the \textquotedblleft speed\textquotedblright \ of star formation quenching and is the quantity which we aim to measure. A smaller $\gamma$ value corresponds to slower quenching, whereas a larger $\gamma$ value corresponds to faster quenching.

In order to determine $\gamma$ for green valley galaxies we based our study on the rest-frame 4000~\AA \ break and the H$\delta$ absorption line (see Section \ref{COSMOS_field} for full details). \citet{Kauffmann2003} have shown that these two spectral indices are sensitive to the galaxy SFH. The 4000 \AA \ break is small for young stellar populations and it becomes larger with older stellar populations, whereas the H$\delta$ absorption is strongest for galaxies that experienced a star formation burst in $0.1-1$~Gyr ago, i.e., when the optical luminosity is dominated by A stars. The 4000~\AA \ break \citep[D$_{n}(4000)$, ][]{Balogh1999} and H$\delta$ absorption line  \citep[H$\delta_A$, ][]{Worthey1997} indices are defined as:
\begin{equation}
  \text{D}_n(4000) = \sum_{\lambda = 4000 \text{\AA}}^{4100 \text{\AA}} F_{\lambda} \left/ \sum_{\lambda=3850 \text{\AA}}^{3950\text{\AA}} F_{\lambda} \right. 
 \label{Dn_4000_indice}
\end{equation} 
and
\begin{equation}
 \text{H}\delta_A = \sum_{\lambda=4083,5 \text{\AA}}^{4122,25 \text{\AA}}\left(1 - \frac{F_{\lambda}}{F_{\lambda,\text{cont}}}\right) d\lambda \,\, ,
 \label{H_delta_A_indice}
\end{equation}
where $F_{\lambda}$ is the flux at wavelength $\lambda$, $F_{\lambda,\text{cont}}$ is the continuum flux, defined by fitting a straight line through the average flux density between 4041.60~\AA \ and 4079.75~\AA, bluewards of the H$\delta$ absorption, and 4128.50~\AA \ and 4161.00~\AA, redwards of it.

We emphasize that this methodology is independent of the absolute value of $t_c$. Since the $\text{D}_n(4000)$ and $H\delta_A$ indices and $\text{NUV}-r$ colour all stabilize after a few Gyr of constant SFR, the methodology is insensitive to the precise duration of that period if it is longer than $\sim 2-3$ Gyr; instead, it determines the average slope in the decline of star formation since the onset of quenching assuming an exponential decline in SFR (we refer the reader to Figures 1 through 3 and Section 2 in \citealt{Nogueira-Cavalcante2018} for more details). This methodology has also yielded quenching time-scales since $z\sim 1$ \citep{Martin2007,Goncalves2012} that are consistent with independent measurements of the growth of the red sequence over the same epoch \citep{Faber2007}. In summary, the value of $t_c$ is marginalised after our choice of measured parameters, and the methodology is only sensitive to $\gamma$. In this work we set $t_c=$ 6 Gyr, for green valley galaxies from COSMOS, as in \citet{Goncalves2012} and \citet{Nogueira-Cavalcante2018}. We test different values of $t_c$ and we verify that this does not change our results.

\section{Galaxy Sample: Cosmos field}\label{galaxy_sample}

We measure star formation quenching time-scales in green valley galaxies at $0.5<z<1.0$ based on a sample of galaxies from the COSMOS\footnote{http://irsa.ipac.caltech.edu/data/COSMOS/} field \citep{Scoville2007}.

\subsection{Photometric Sample}

We adopt the same procedure as presented in \citet{Goncalves2012} and \citet{Nogueira-Cavalcante2018}. We select galaxies with detections in all five Canada-France-Hawaii Telescope Legacy Survey (CFHTLS) bands ($u,g,r,i,z$). This is a necessary condition to properly constrain the equivalent SDSS $r$-band and, especially, the GALEX Near-Ultraviolet-band (NUV-band) rest-frame magnitudes of the galaxies, since the rest-frame NUV emission at $0.5<z<1.0$ is shifted to the $u$ and $g$ CFHTLS bands. The NUV$-r$ colour efficiently separates the blue and red populations due to its increased dynamic range when compared to optical colours \citep{Wyder2007}. We use the $K$-correct code \citep[version 4.2;][]{Blanton2007} to calculate the rest-frame magnitudes, given the photometric redshifts determined by the CFHTLS consortium \citep{Ilbert2006,Coupon2009}.

\subsubsection{Extinction Correction}\label{extinction_correction}

Most green valley galaxies are dust-obscured star-forming galaxies, shifted towards redder colours and into the green valley due to significant dust extinction. In fact, \citet{Sodre2013} have shown that the reddest galaxies in the local universe are star-forming obscured disks. At intermediate redshifts, \citet{Goncalves2012} demonstrated that up to 65\% of all bright green valley galaxies can be dusty star-forming galaxies. In order to remove these galaxies from the green valley we use the extinction correction model from \citet{Salim2009}. Following this approach, the extinction in the Far-Ultraviolet (FUV) magnitude is given by
\begin{equation}
A_{FUV} = 3.68(\text{FUV}-\text{NUV})+0.29 \,\, 
\end{equation}
and the extinction correction in other filters can be determined using the following relation:
\begin{equation}
 A_{\lambda}\propto \lambda^{-0.7} \,\, .
\end{equation}

\begin{figure}
\includegraphics[width=\columnwidth]{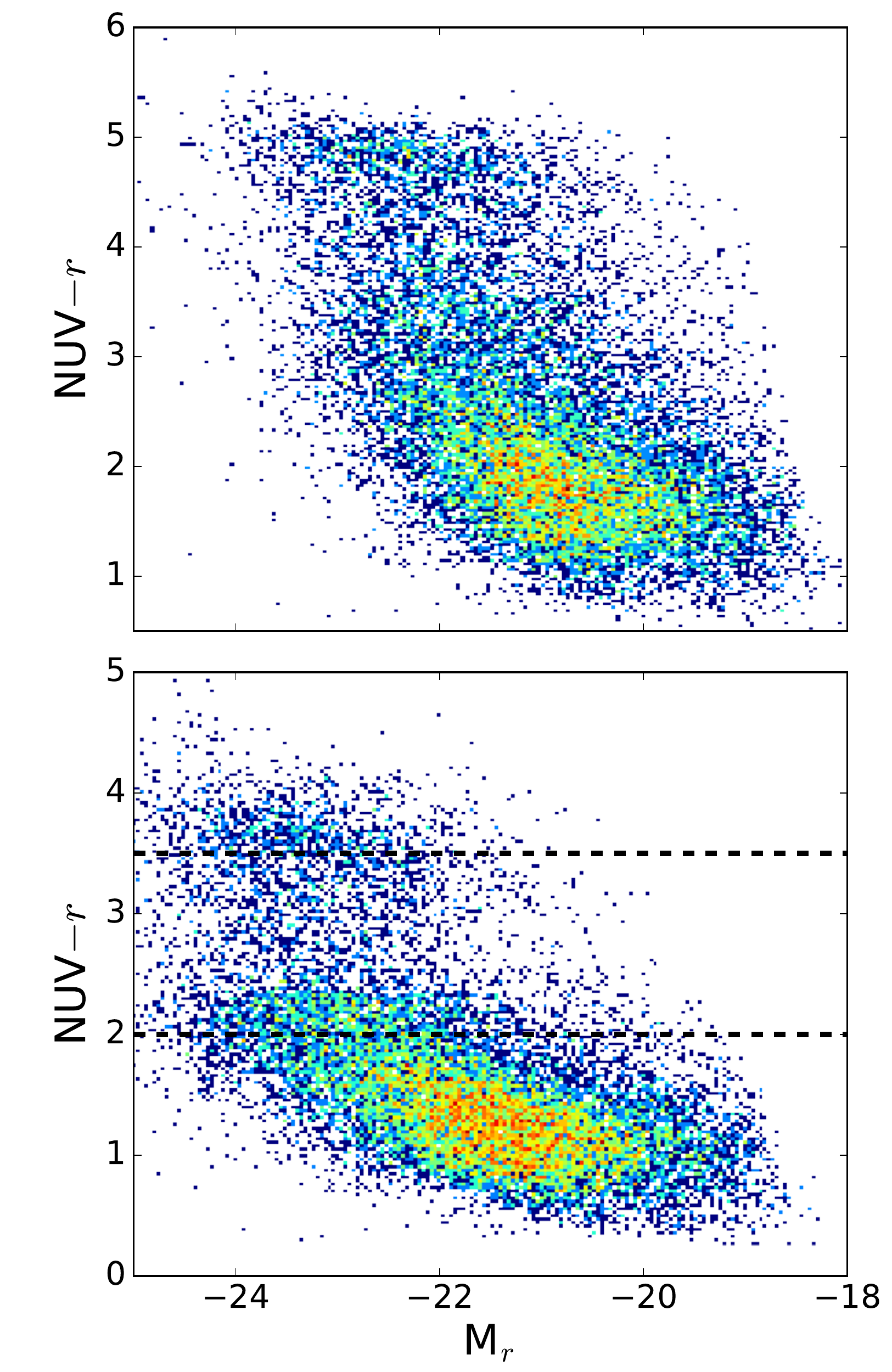}
\caption{The $\text{NUV}-r \times \text{M}_r$ colour-magnitude diagrams of CFHTLS galaxies, where galaxy magnitudes are uncorrected (top panel) and corrected (bottom panel) by internal dust extinction. The dashed black lines in bottom panel define the green valley region.}
\label{color_magnitude_diagram_CFHTLS}
\end{figure}

We note that this procedure works well for dusty star-forming galaxies, but over-corrects the colours of dust-poor galaxies \citep{Salim2009}. Figure \ref{color_magnitude_diagram_CFHTLS} shows the colour-magnitude diagram of the $\sim19000$ CFHTLS galaxies with and without correction for dust extinction. As in \citet{Goncalves2012} and \citet{Nogueira-Cavalcante2018}, we define our green valley galaxy sample by selecting objects with $2.0<$ NUV$-r<3.5$ in the dust-corrected CMD in an effort to mitigate the contamination of the green valley by dusty star-forming galaxies. Once we have minimised the presence of dusty star-forming galaxies from the green valley we make the reasonable assumption that the observed colours of the green valley galaxy sample are effectively unaltered by dust. For this reason, we use the observed colours for the rest of the analysis. From the Hubble Space Telescope (HST) images, we note that these dusty star-forming galaxies are mostly disks ($\sim75\%$), with few elliptical galaxies ($\sim15\%$) and $\sim10\%$ currently undergoing interactions. According to our compactness definition (Eq. \ref{compactness_definition_equation}, Section \ref{compactness_definition}), $\sim98\%$ are normal-sized galaxies whereas $\sim2\%$ are compact ones.

\subsection{Spectroscopic Sample}

We exploit the zCOSMOS DR3 spectroscopic survey \citep{Lilly2007} to measure the D$_n(4000)$ and H$\delta_A$ spectral indices of our green valley galaxies. This survey comprises $\sim$20,000 spectra of galaxies in the COSMOS field with magnitudes down to $i=22.5$, taken with the VIsible Multi-Object Spectrograph (VIMOS) located on the ESO Very Large Telescope (VLT). The spectral coverage of these spectra is from 5500\AA \ to 9700\AA, with a resolution of $R \sim 600$. This spectral coverage is ideal to measure the H$\delta_A$ and D$_n$(4000) indices within our redshift range of interest $(z\sim0.5-1)$.

\subsection{UltraVISTA Catalogue}

To define our sample of compact, massive green valley galaxies, we obtain stellar masses from \citet{Muzzin2013}, whose catalogue provides photometry in 30 bands, from UV to near-IR, including data from GALEX \citep{Martin2005}, CFHT/Subaru \citep{Capak2007}, UltraVISTA \citep{McCracken2012} and S-COSMOS \citep{Sanders2007}. The authors determined the stellar population parameters (including stellar mass) by fitting the galaxy spectral energy distribution (SED) through the FAST code \citep[Fitting and Assessment of Synthetic Templates, ][]{Kriek2009}. This code uses the \citet{Bruzual2003} models, assuming solar metallicity, \citet{Chabrier2003} initial mass function and the \citet{Calzetti2000} dust extinction law. \citet{Muzzin2013} compared their estimated stellar masses with those from the NEWFIRM Medium Band Survey \citep[NMBS, ][]{Whitaker2011} catalogue, which is derived using the EAZY \citep{Brammer2008} and P\'EGASE \citep{Fioc1999} codes, \citet{Maraston2005} stellar population models and \citet{Kroupa2001} initial mass function. The agreement of the two catalogues is good, within the range of 0.3-0.4 dex.

\subsection{Compactness Definition}\label{compactness_definition}

A variety of definitions for compactness exists in the literature \citep[e.g., ][]{Carollo2013, Quilis2013, Dokkum2015}. In order to better understand the role of galaxy compactness in quenching star formation in green valley galaxies, for this work, we apply the compactness definitions proposed by \citet{Wel2014}, where these authors define two categories for compact galaxies:

\begin{equation}\label{compactness_definition_equation}
A = \frac{R_{\text{eff}}}{\left( M_\star/10^{11}\,M_{\odot} \right)^{0.75}} \,\, ,
\end{equation}
where $R_{\text{eff}}$, in units of kpc, is the effective radius (containing half of the total flux), $M_\star$ is the galaxy stellar mass, and $A<1.5$, $1.5<A<2.5$ and $A>2.5$, in units of kpc, define \textit{ultra-compact}, \textit{compact} and \textit{normal-sized} galaxies, respectively. 

\begin{figure}
\includegraphics[width=\columnwidth]{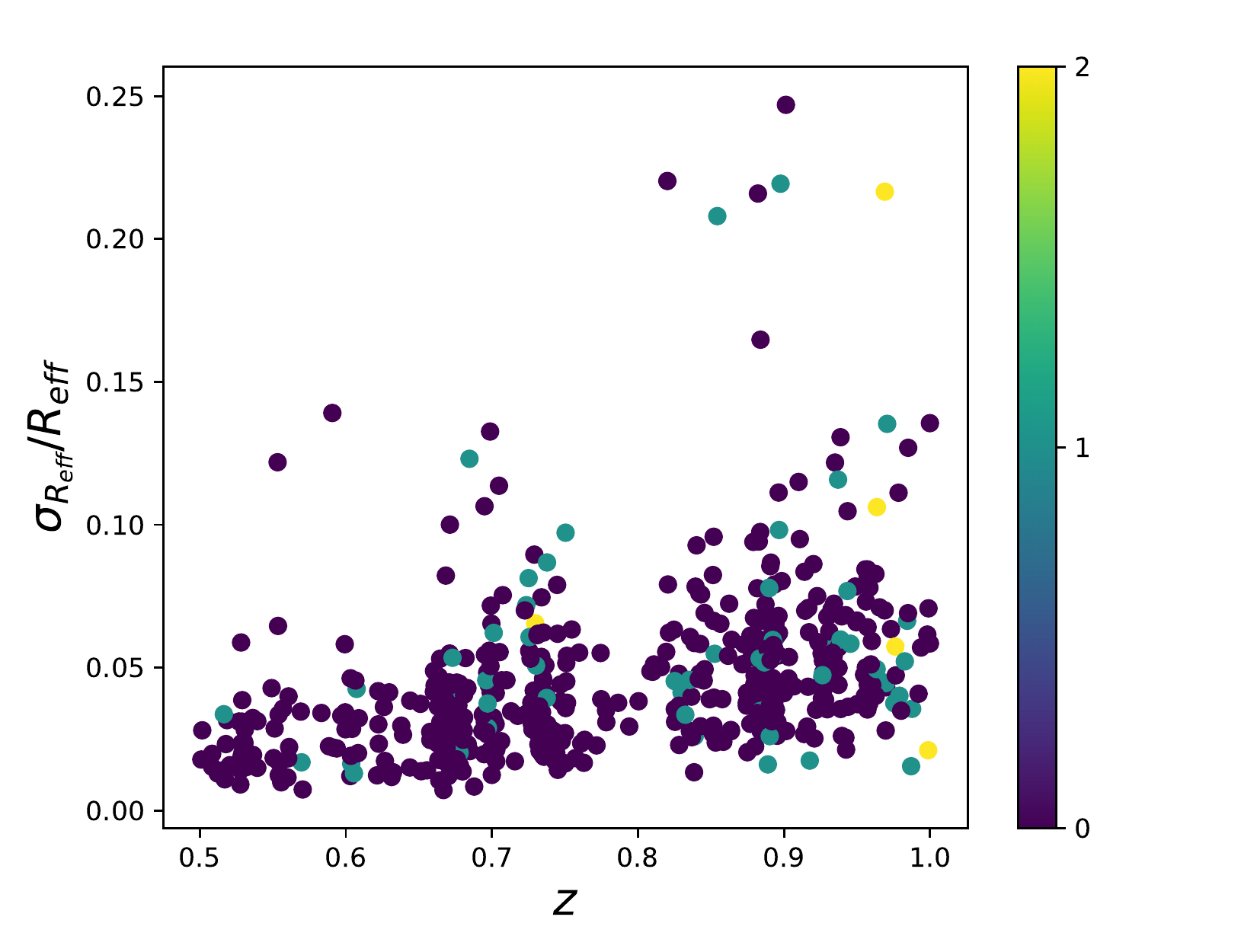} 
\caption{Error of R$_eff$ over R$_eff$ ($\sigma_{R_{eff}}/R_{eff}$) as a function of redshift. The colours represent the compactness of galaxies from our sample ($0=\text{normal-sized}, 1=\text{compact} \text{ and } 2=\text{ultra-compact}$).}
\label{Error_Reff_over_Reff_as_a_function_of_redshift}
\end{figure}

We estimate the effective radius $R_{\text{eff}}$ using the following conversion:
\begin{equation}
 R_{\text{eff}} = D_A(z) \times R_{GIM2D}
 \label{effective_radius}
\end{equation}
where $D_A(z)$ is the cosmological angular distance and $R_{GIM2D}$ is the PSF-convolved half-light radius of the object. This information is taken from the Zurich Estimator of Structure Types (ZEST) catalogue, which comprises, among others, structural parameters, like Sersic index and angular sizes, measured from \citet{Sargent2007}. In Figure \ref{Error_Reff_over_Reff_as_a_function_of_redshift} we show the evolution of error of R$_eff$ divided by R$_eff$ ($\sigma_{R_{eff}}/R_{eff}$) with redshift, showing that this relative error is almost constant in the entire redshift range, except for few galaxies whose relative errors are greater than 0.1.

\subsection{Environment Catalogue}\label{environment_catalog}

External secular processes driven by environment play an important role in quenching star formation in galaxies, specially in scales $<1$ Mpc \citep{Kauffmann2003}. Processes such as harassment, strangulation and ram pressure stripping can remove and/or heat molecular gas in galaxies, stopping and/or preventing star formation \citep{Gunn1972, Larson1980, Farouki1981, Merritt1983, Moore1998, Abadi1999, Balogh1999, Poggianti1999, Balogh2000, Coil2008, Peng2015, Smethurst2017}. 

In order to remove possible environmental dependencies we must exclude from our data galaxies that potentially reside in dense environments. We obtain the local density of green valley galaxies from \citet{Darvish2015}. In that work the authors derived local environmental densities using the weighted version of the Voronoi tessellation method \citep[weighted adaptive kernel, ][]{Darvish2014} in a $K_{s}<24$ mag sample in the COSMOS field \citep{Capak2007}, with photometric redshifts from the UltraVISTA catalogue \citep{McCracken2012, Ilbert2013} at $0.1<z<3.0$. The quantity used to express the environment of galaxies is the \textit{overdensity} $(1+\delta)$. Briefly, in order to estimate the local environment \citet{Darvish2015} built a series of redshift slices ($z$-slice) on their sample. The widths of the slices are obtained from the probability distribution function (PDF) of the galaxies: for each slice, the width is defined as twice the median redshift uncertainty of the galaxies at each redshift. The overdensity in each galaxy $(1+\delta_i)$ is defined as the local surface density divided by the median local surface density within each $z$-slice, at the position of the $i$th galaxy.

\subsection{Galaxy Sample Selection}\label{sample_selection}

Based on our parent sample of green valley galaxies in the redshift range $0.5<z<1.0$ we define sub-samples of normal-sized green valley galaxies and both compact and ultra-compact green nuggets, following Equation \ref{compactness_definition_equation}. In agreement with the analysis and compactness definition of \citet{Wel2014}, we restrict our sample of study to galaxies with stellar mass M$_{\star}>10^{10.7}$ M$_{\odot}$.

\begin{figure}
\includegraphics[width=\columnwidth]{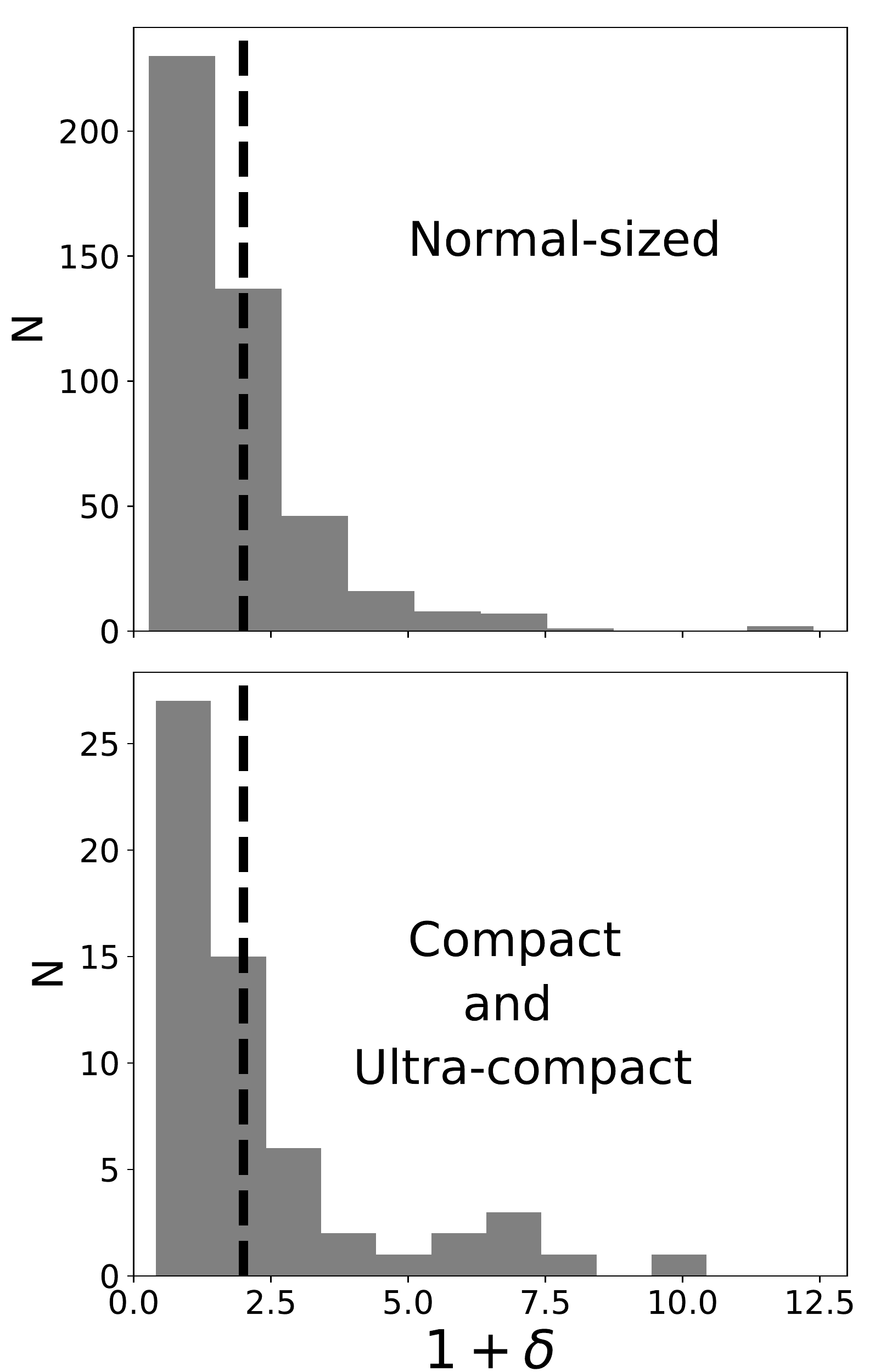} 
\caption{Overdensity distribution for normal-sized (top) and compact and ultra-compact (bottom) green valley galaxies which have zCOSMOS spectra. The vertical dashed black line in each panel delimits the overdensity limit for low-to-moderate environments ($1+\delta < 2.0$).}
\label{overdensity_distribution_normal_sized_and_compact_green_valley_galaxies}
\end{figure}

Figure \ref{overdensity_distribution_normal_sized_and_compact_green_valley_galaxies} shows the overdensity ($1+\delta$) distribution for normal-sized, compact and ultra-compact green valley galaxies. According to \citet{Darvish2015} high overdensities are defined as $1+\delta > 3.0$ and low overdensities as $1+\delta < 1.5$. In an effort to simultaneously remove the influence of environment and maintain a reasonable number of green valley galaxies in our sample, we exclude from our analysis green valley galaxies in environments characterised as high overdensities and select green valley galaxies with low-to-moderate environments (i.e.,  $1+\delta<2.0$). Our partial sample is divided into 311 normal-sized, 34 compact and 5 ultra-compact green valley galaxies. 

We have checked the HST galaxy images of the compact and ultra-compact galaxies and note that some of these galaxies have a nearby companion. To avoid contamination from nearby companions, whose light may alter the spectrum associated to our target, we exclude from our sample galaxies with a companion within 5 arcsecs, from the HST images provided with the zCOSMOS galaxy spectra. This angular size corresponds a physical scale of $\sim30-40$ kpc. The resulting final numbers of normal-sized, compact and ultra-compact galaxies are shown in Table \ref{table_quenching_values_and_median_color_compactness_study}. 

\begin{table}
\begin{center}
\begin{tabular}{|c|c|c|c|c|c|}
\hline
Compactness & Number of & $\langle$NUV$-r\rangle$ & $\gamma$          \\
            &  galaxies &                         &  [Gyr$^{-1}$]      \\
\hline \hline
Normal-sized   & 284 & 4.0 $\pm$ 0.5 & 2.7 $\pm$ 0.6  \\
($A>2.5$)      &     &               &                \\
Compact        & 23  & 4.3 $\pm$ 0.4 & 5.0 $\pm$ 3.2  \\
($1.5<A<2.5$)  &     &               &                \\
Ultra-compact  & 3   & 4.3 $\pm$ 0.3 & 11.0 $\pm$ 3.0 \\
($A<1.5$)      &     &               &                \\
\hline
\end{tabular}
\caption{Number of galaxies, median colour values and quenching index for each compactness type.}
\label{table_quenching_values_and_median_color_compactness_study}
\end{center}
\end{table}

\section{Simulated sample: Illustris project}\label{illustris_simulation_description}

We use the public release of the Illustris simulation \citep{Vogelsberger2014, Nelson2015}, readily accessible through an application programming interface (API) web, where abundant data on individual sub-halos can be downloaded. Within the context of simulations, sub-halos are gravitationally-bound sub-structures made of dark matter and baryons. For our purposes we call these sub-structures {\it galaxies}. 

\begin{figure}
  \includegraphics[width=\columnwidth]{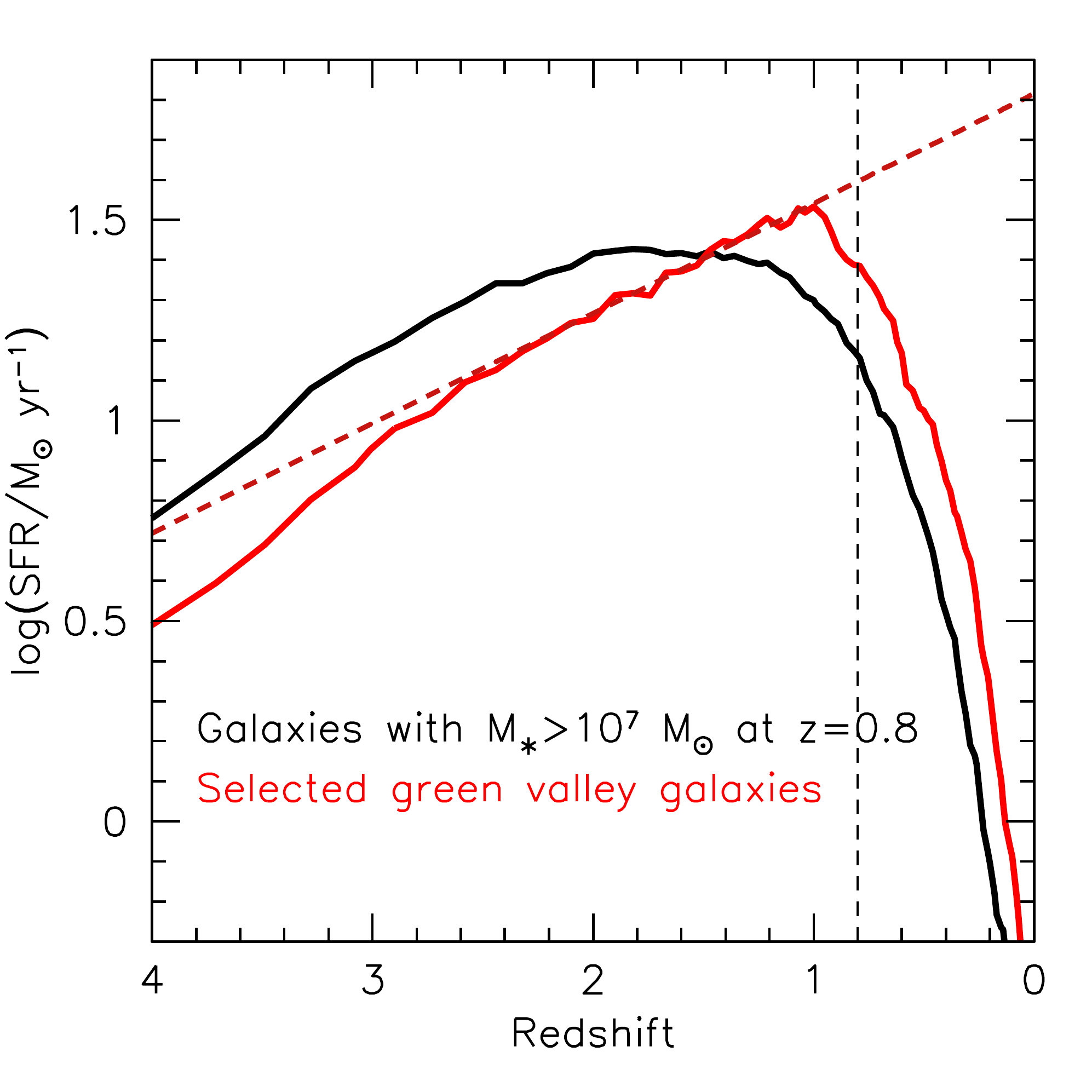}
  \caption{SFHs of Illustris simulated galaxies. Black curve represents the median SFH for the whole Illustris galaxy sample (1305 galaxies), with masses $M_{\star}~>~10^{10.7} M_{\odot}$ at $z=0.8$ (black vertical line), while the red curve represents the median SFH of green valley sub-sample (331 galaxies), selected as explained in Section \ref{illustris_simulation_description}. The dashed red line is the best fit to the red SFH in the redshift interval between $z=2$ and $z=1$.}
  \label{GVselect}
\end{figure}

Our goal is to match both the simulated and observed galaxy samples. First, simulated galaxies are constrained to have stellar masses larger than $M_{\star}~>~10^{10.7} M_{\odot}$ \citep[condition from][]{Wel2014}. We select 1305 Illustris galaxies with these conditions. Additionally, simulated galaxies should reach the green valley phase by at least $z=0.8$. Figure \ref{GVselect} shows the median SFHs for both the selected green valley sample (331 galaxies in red) and the whole sample (1305 galaxies in black). Typically, simulated massive galaxies live an active star formation phase from $z=4$ to $z=2$, followed by a drastic quenching process, which drives them to a quiescence state. To select galaxies in the green valley phase we require them to initiate their SF quenching not earlier than $z=1$. As discussed later in Section \ref{conclusions_and_discussion}, our simulated galaxies typically quench in 1 Gyr, therefore they are expected to remain in the green valley phase from $z=1$ to $z=0.8$. The black curve in Figure \ref{GVselect} shows that the majority of massive galaxies at $z=0.8$ should have left the green valley long ago, as they started quenching at $z\sim2$, 3.5 Gyrs earlier than $z=0.8$.

Our approach is to use the shape of the individual SFH curves to select galaxies starting their SF quenching at $z\sim1$. The SFH shape can be outlined by two sections, before and after $z\sim1$. We perform a linear fit to the SFH in the redshift interval of $z=1-2$ in order to constraint the pre-quenching phase (red dashed line in Figure \ref{GVselect}). We impose the following three conditions: (1) the linear fit (from $z=2$ to $z=1$) must show a positive slope, in order to exclude quenching earlier than $z=1$; (2) the individual SFHs must peak around $z=1$ or slightly later. This imposes that the SFH must be close to the linear fit value at $z\sim1$, allowing for a difference of up to 0.18 dex (average rms value of the linear fit); and (3) the SFH slope at $z=0.5-0.8$ interval must be negative, to exclude temporary SF fluctuations. With this procedure we select a final sample of 331 simulated galaxies. Their median SFH is shown in Figure \ref{GVselect} (red curve). 

The normal-sized, compact and ultra-compact sub-samples are defined following Eq. \ref{compactness_definition_equation}. We use the projected effective radius ($R_{eff,2D}$), which, according to \citet{Wellons2015}, is 75\% of the 3D radius ($R_{eff,3D}$). The numbers of green valley galaxies in the Illustris sub-samples are shown in Table \ref{table_compacteness_and_number_of_galaxies_from_Illustris_sample}.

\begin{table}
\begin{center}
\begin{tabular}{|l|c|c|}
\hline

Compactness & Number of & $\langle \gamma \rangle$          \\
            &  galaxies & [Gyr$^{-1}$]      \\

\hline \hline
Normal-sized ($A>2.5$) & 315 & 0.35\\
Compact ($1.5<A<2.5$) & 14 & 1.29\\
Ultra-compact ($A>2.5$) & 2 & 2.23\\
\hline
\end{tabular}
\caption{Compactness definition, number of green valley galaxies and $\gamma$ values from Illustris galaxy sample.}
\label{table_compacteness_and_number_of_galaxies_from_Illustris_sample}
\end{center}
\end{table}

\section{Results}\label{results}

\subsection{COSMOS Field}\label{COSMOS_field}

\begin{figure*}
\begin{center}
\includegraphics[width=\textwidth]{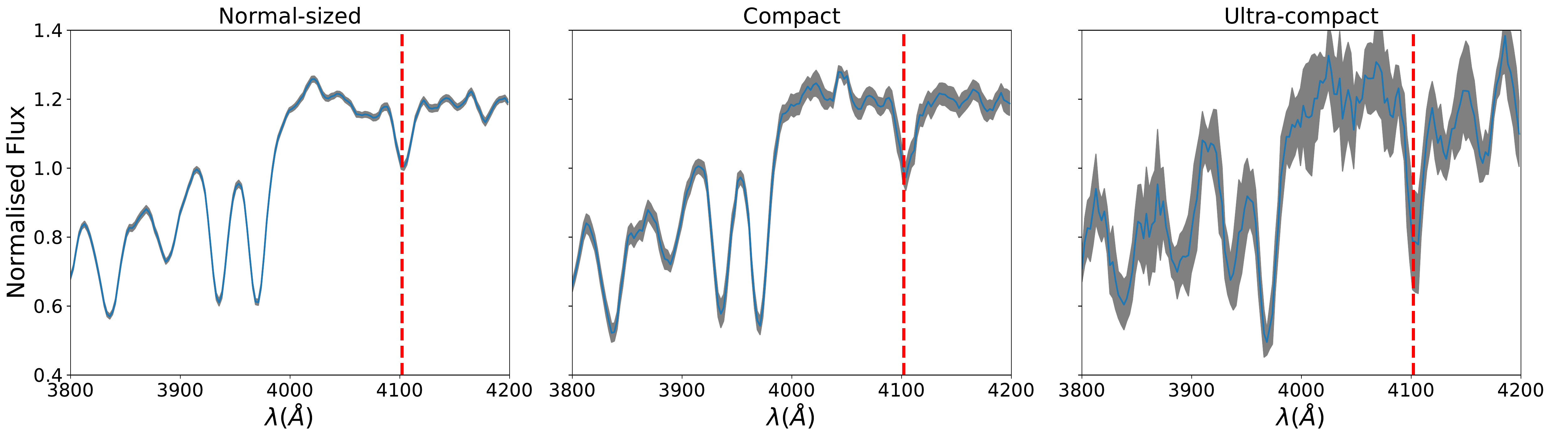}
\includegraphics[width=\textwidth]{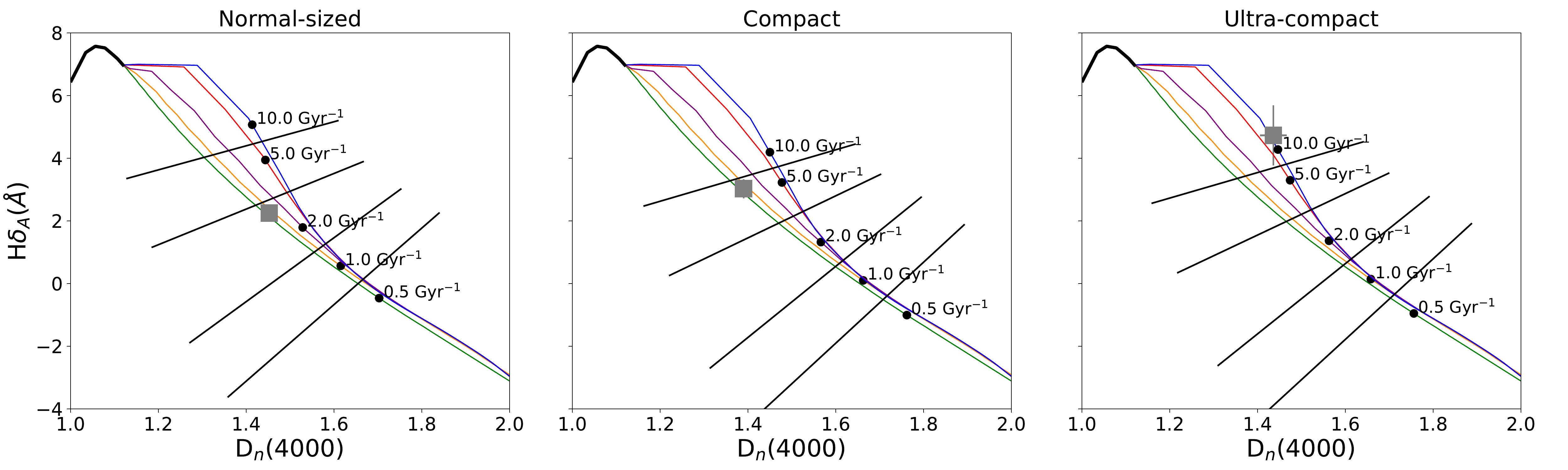}
\caption{\textit{\textbf{Top:}} Coadded zCOSMOS spectra (blue lines) and the standard deviation (grey region) for each compactness group. The red vertical dashed line indicate the H$_{\delta}$ absorption line. \textit{\textbf{Bottom:}} H$\delta_A \times$ D$_n(4000)$ planes for each compactness group, where the exponential index $\gamma$ is estimated. The gray square in each plane is the spectral indices point, measured from the coadded spectra, shown in the top panels. The coloured curves are the spectral indices evolution predicted from \citet{Bruzual2003} models, using the Eqs. \ref{star_formation_history_1} and \ref{star_formation_history_2}. To reproduce these curves we consider five different SFH models: $\gamma$ [Gyr$^{-1}$] = 0.5 (green), 1.0 (orange), 2.0 (purple), 5.0 (red) and 20.0 (blue). The black dots represent the D$_n(4000)$ and H$\delta_A$ values, for each SFH model, when it reaches the average NUV$-r$ colour specified in the Table \ref{table_quenching_values_and_median_color_compactness_study}. The straight lines represent H$\delta_A$ and D$_n(4000)$ points which are associated with the geometric average between two consecutive black dots. This procedure allows us to interpolate the entire plane, associating a set of galaxy observable (NUV$-r$, D$_n(4000)$ and H$\delta_A$) with a single $\gamma$ value.}
\label{coadded_galaxy_spectra_and_h_delta_vs_dn4000_for_compactness_study}
\end{center}
\end{figure*}

The signal-to-noise ratio (S/N) of individual zCOSMOS spectra is not sufficient to measure the spectral indices with reasonable uncertainties to distinguish between different SFHs. Based on the need for high S/N to determine star formation quenching time-scales, we produce a coadded (stacked) spectrum for each compactness sub-sample: normal-size, compact and ultra-compact. The resulting green valley coadded spectra for each compactness classification are shown in the top panels of Figure \ref{coadded_galaxy_spectra_and_h_delta_vs_dn4000_for_compactness_study}.

\begin{figure}
\begin{center}
\includegraphics[width=\columnwidth]{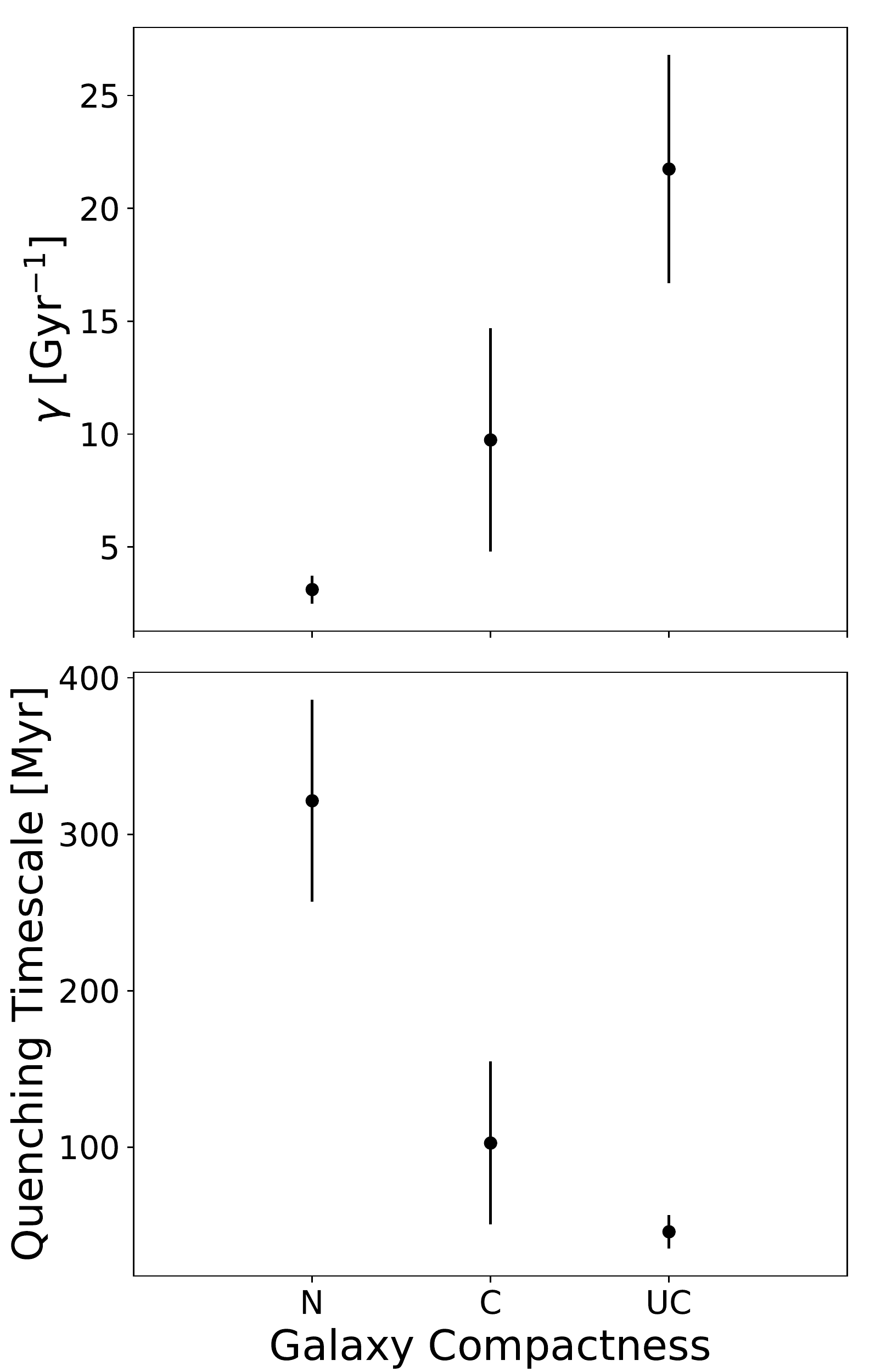}
\caption{The exponential index ($\gamma$) values (top) for the green valley galaxies classified as normal-sized (N), compact (C) and ultra-compact (UC) and quenching time-scales (bottom) when the initial star formation rate decreases to $\sim$37\% of the initial value ($1/\gamma$).}
\label{quenching_index_and_quenching_timescale_as_a_function_of_compactness}
\end{center}
\end{figure}

The bottom panels of Figure \ref{coadded_galaxy_spectra_and_h_delta_vs_dn4000_for_compactness_study} show $-$ for the mean NUV$-r$ colours of each of the compactness samples $-$ H$\delta_A$ $-$ D$_n$(4000) diagrams with the different modelled behaviours expected for SFHs with varying quenching indices ($\gamma$); the H$\delta_A$ and D$_n$(4000) values are shown for each compactness sample. Based on the location of the sub-sample's H$\delta_A$-D$_n$(4000) data-point with respect to the green valley positions within model curves, we determine empirically the star formation quenching indices ($\gamma$) for each compactness sample. For more details on the procedure to determine $\gamma$ from the H$\delta_A$ $-$ D$_n$(4000) diagram we refer the reader to \citet{Nogueira-Cavalcante2018}. Briefly, we adopt four perpendicular bisectors in each diagram, that represent the geometric average between two models predicted by the models discussed in Section \ref{methodology} (black dots in the bottom panels of Figure \ref{coadded_galaxy_spectra_and_h_delta_vs_dn4000_for_compactness_study}). Afterwards, we interpolate the D$_n$(4000) $\times$ H$\delta_A$ plane, associating each coordinate of the plane to a single $\gamma$ value from a set of three observables: NUV$-r$, D$_n$(4000), and H$\delta_A$. This procedure is an attempt to take into account the diversity of star formation histories of the galaxies. As in \citet{Nogueira-Cavalcante2018} we perform 1000 times the estimation of $\gamma$ index for each compactness group, taking into account the errors of galaxy colours and spectral indices ($\left\langle \text{NUV}-r \right\rangle \pm \sigma_{\left\langle\text{NUV}-r\right\rangle}$, D$_{n}(4000)\pm \sigma_{\text{D}_{n}(4000)}$ and H$\delta_A\pm \sigma_{\text{H}\delta_A}$). We obtain the errors of the spectral indices from the flux uncertainties, from coadded galaxy spectra (gray regions in Figure  \ref{coadded_galaxy_spectra_and_h_delta_vs_dn4000_for_compactness_study}). For the errors in NUV$-r$ colours we consider the standard deviation of NUV$-r$ colours of green valley galaxies in each compactness group. Therefore, with this procedure, we obtain a distribution of 1000 $\gamma$ values for each compactness group. We set up a final $\gamma$ and its error by the average and the standard deviation of the distribution of $\gamma$ values, in each compactness group. The estimated values of $\gamma$ for each green valley compactness group and the average NUV$-r$ colours of the galaxies in each compactness bin are also shown in Table \ref{table_quenching_values_and_median_color_compactness_study}. Figure \ref{quenching_index_and_quenching_timescale_as_a_function_of_compactness} shows a direct comparison of star formation quenching indices and time-scales for all 3 bins of compactness. From this we observe that the quenching time-scales of green nuggets (approximately 100 Myrs and 160 Myrs for ultra-compact and compact green valley galaxies, respectively) are relatively much shorter than that of normal-sized green valley galaxies ($\sim 370$ Myrs). 

\subsection{Illustris Simulation}\label{illustris_simulation_results}

\begin{figure}
  \includegraphics[width=\columnwidth]{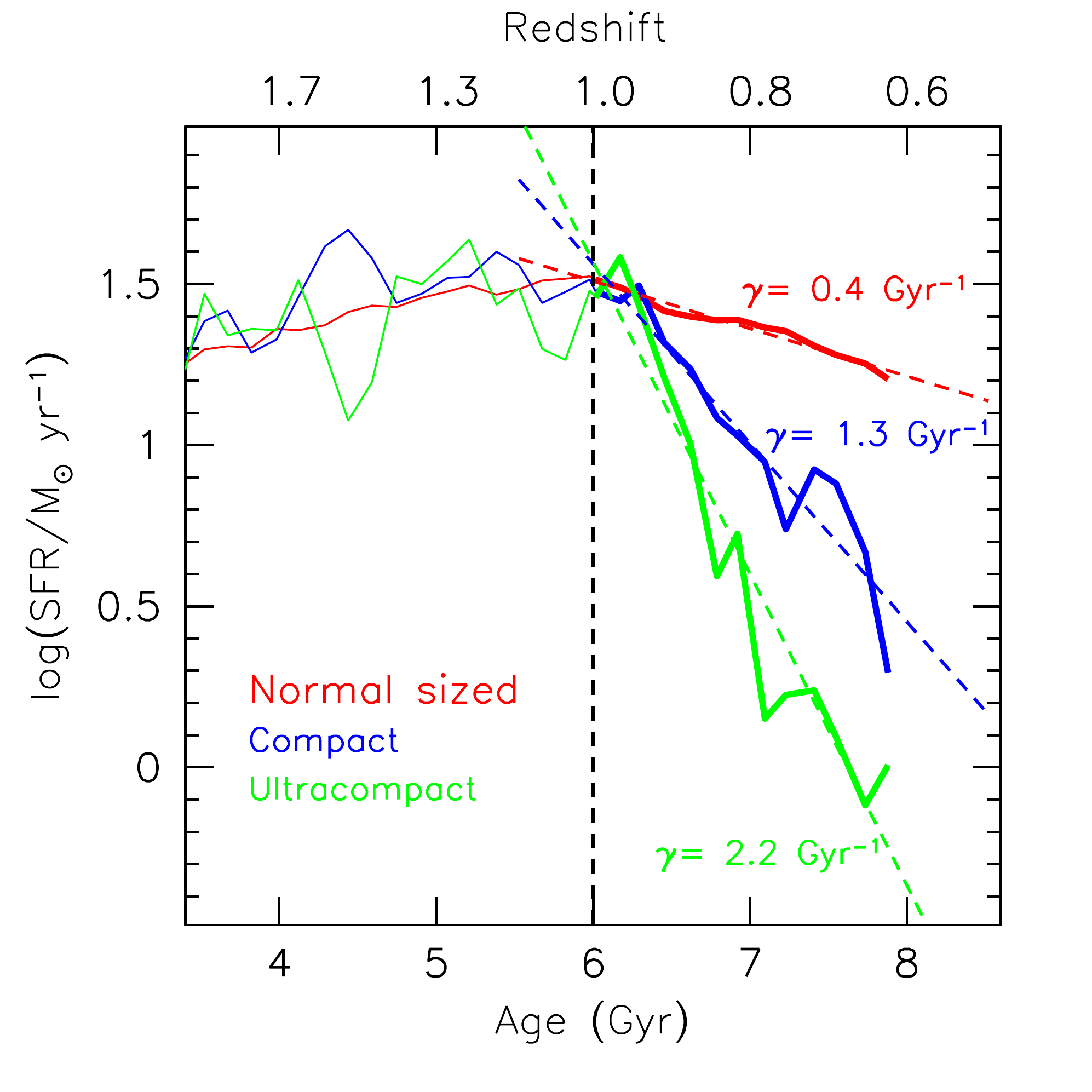}
  \caption{The median star formation history (SFH) of the normal-sized (red), compact (blue) and ultra-compact (green) green valley galaxies from Illustris data. The x-axes correspond to the age of the universe (bottom) and the redshift (top). Tilted dashed lines show the least square fits to the SFR decaying phase along 2 Gyr after the SF quenching onset at $z\sim1$ ($\sim6$ Gyrs). Clearly, the $\gamma$ values increase with the compactness, in qualitative (not quantitative) agreement with Figure  \ref{quenching_index_and_quenching_timescale_as_a_function_of_compactness}}.
  \label{sfh_illustris_median_evolution}
\end{figure}

\begin{figure}
\includegraphics[width=\columnwidth]{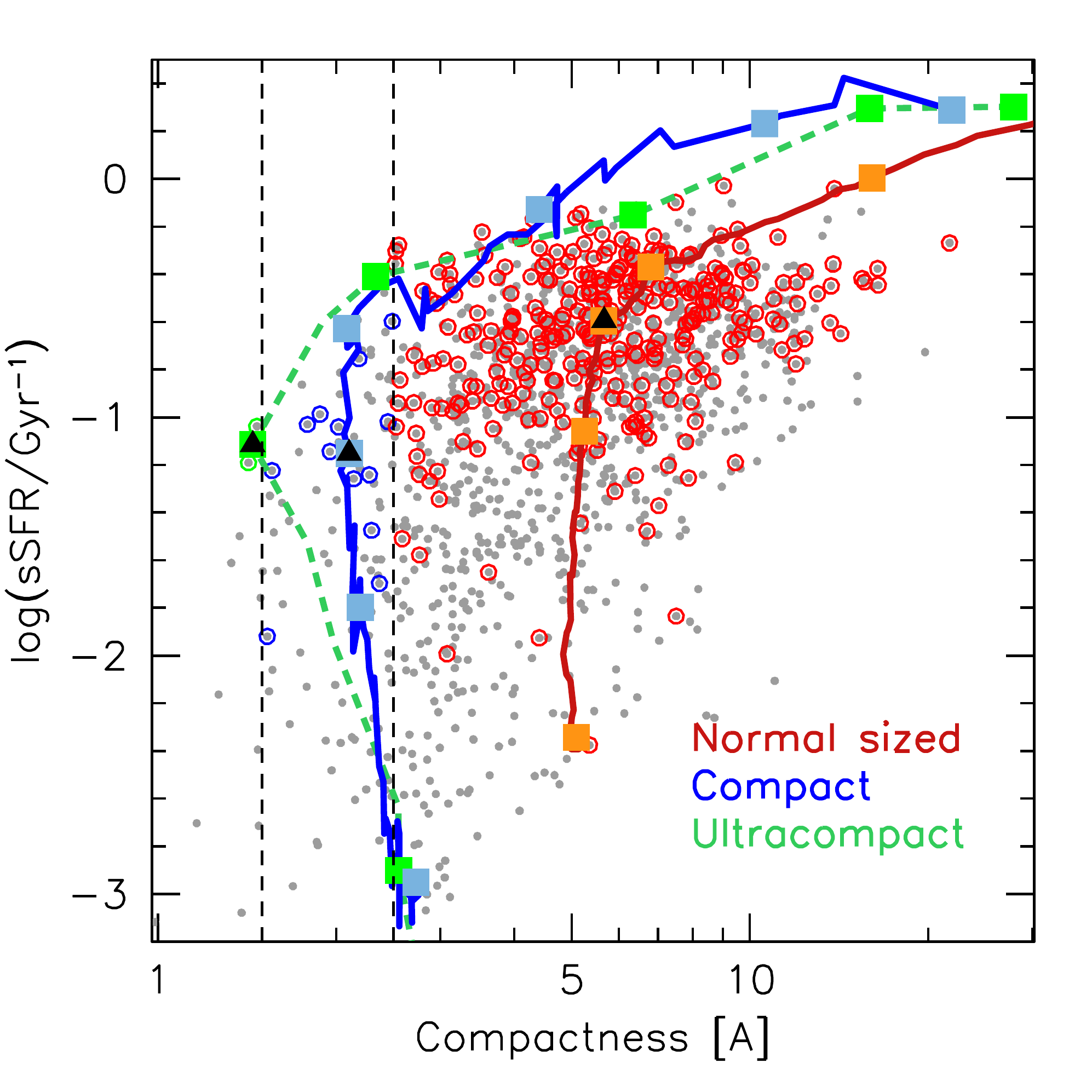}
\caption{Evolution from $z=4$ to $z=0$ of compactness ($A$) and specific star formation rate (sSFR) median values for the Illustris compact (blue) and normal-sized (red) green valley sub-samples. The ultra-compact sub-sample, with only two galaxies, is represented by an average low resolution track (dashed green). Note that compact (normal-sized) galaxies have low (high) $A$ values and vertical dotted lines mark the limits of the three compactness sub-samples. Background grey dots are all M$_{\star}>10^{10.7}$~M$_{\odot}$ galaxies at $z=0.8$, with green valley members highlighted by open coloured circles. Evolutionary tracks show marks (coloured squares) in some selected redshifts, $z=4, 3, 2, 1, 0.8, 0.5 \text{ and } 0$. Their redshift is easy to identify as $z=0.8$ is highlighted by a black triangle and redshift values increase with sSFR.}
\label{plot_evol_logA}
\end{figure}

In this section we track the evolution of several parameters from $z=4$ to $z=0$ for our Illustris sub-samples of (compact and normal-sized) simulated galaxies in terms of \textit{median} and \textit{individual} perspectives. In the first case, we track median values for each sub-sample and in the second case we select a representative galaxy from each sub-sample. To simplify our results, in some cases we overlook the ultra-compact galaxies, because this sub-sample contains only two galaxies that can readily fit as extreme cases of the compact subsample. Nevertheless, ultra-compact galaxies are shown separately in Figs. \ref{sfh_illustris_median_evolution} and \ref{plot_evol_logA}.

\begin{figure}
\includegraphics[width=\columnwidth]{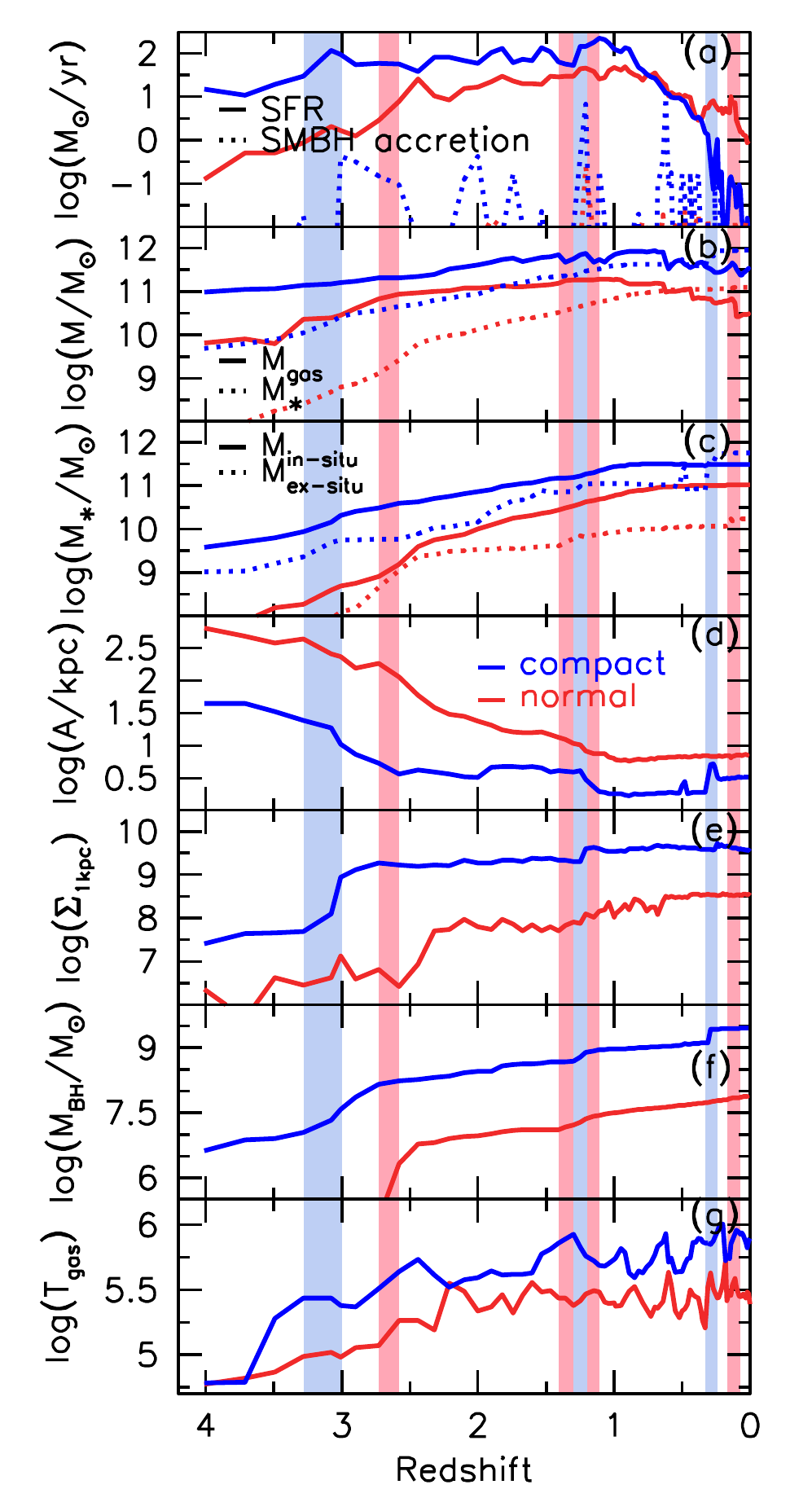}
\caption{Parameter evolution, from $z=4$ to $z=0$, of compact (blue) and normal-sized (red) Illustris green valley galaxies: (a) SFR (full line) and central super massive black hole (SMBH) accretion (dashed); (b) stellar (dashed) and gas (full line) mass; (c) in-situ (continuous) and ex-situ (dashed) stellar masses; (d) compactness parameter ($A$), in which higher values correspond to a more diffuse structure; (e) $\Sigma_{1kpc}$ (stellar mass density in the central 1 kpc); (f) SMBH mass and (g) circumgalactic gas temperature evolution. The shaded vertical bands (bluish/reddish for the compact/normal examples) mark the most influential mergers.}
\label{evolution_of_parameters_individual_evolution}
\end{figure}

\subsubsection{Median evolution of simulated normal-sized and compact galaxies}\label{median_evolution}

Figure \ref{sfh_illustris_median_evolution} shows a detailed version of the red curve in Figure \ref{GVselect} with differences between the three compactness sub-samples highlighted. Median SFHs are presented for the normal-sized (red), compact (blue) and ultra-compact (green) sub-samples.

We decrease, at age = 6 Gyrs, in $\sim0.33$ dex the log(SFR) both for compact and ultra-compact green valley Illustris galaxies in order to compare directly with the log(SFR) of normal-sized ones. We also perform a linear fit of the decaying-phase of the SFH curves, from 6 to 8 Gyr (at $z\sim0.6-1$), to derive the quenching indices from the linear fit slope. Clearly, from Figure \ref{sfh_illustris_median_evolution}, ultra-compact and compact green valley galaxies ($\gamma \text{ [Gyr}^{-1}\text{]} = 2.2 \text { and } 1.3$, respectively) quench faster than their normal-sized counterparts ($\gamma \text{ [Gyr}^{-1}\text{]} = 0.35$). The corresponding quenching time-scales ($1/\gamma$ [Gyr]) are 0.45, 0.78 and 2.86, respectively for ultra-compact, compact and normal-sized galaxies. These results are in qualitative agreement with our results using real green valley galaxies in the COSMOS field, although the quenching time-scales in Illustris galaxies are much longer than those found in real green valley galaxies. We discuss the implication of these findings in Section \ref{conclusions_and_discussion}.

Figure \ref{plot_evol_logA} shows the evolution of the compactness ($A$, defined in Equation \ref{compactness_definition_equation}) and the specific star formation rate (sSFR) for normal-sized, compact and ultra-compact Illustris galaxies from $z=4$ to $z=0$. The normal-sized and compact evolutionary tracks correspond to median values, while the ultra-compact track, shown with low temporal resolution, is the average of the only two subsample galaxies. Background grey dots represent the full sample of M$_{\star}~>~10^{10.7}$~M$_{\odot}$ galaxies at $z=0.8$, with the selected compactness sub-samples highlighted by colored circles. Vertical dotted lines mark the separation between the compactness sub-classes. Colored squares mark different redshifts ($z=4, 3, 2, 1, 0.8, 0.5 \text{ and } 0$). Identification is straightforward as $z=0.8$ are distinguished with black triangles and redshift values increase with sSFR. We emphasize that green valley galaxies at $z=0.8$ were not selected according to a particular sSFR cut, but to the SFH shape, as explained in Section \ref{illustris_simulation_description}. Therefore, the coexistence of green valley (red, blue and green circles) and non green valley (grey dots) galaxies in Figure \ref{plot_evol_logA} turns up to reveal the diverse evolutionary phases in which the massive galaxies at $z=0.8$ are found.

Tracks in Figure \ref{plot_evol_logA} show that all green valley galaxies share similarly high sSFR and low compactness values at $z=4$. As redshift decreases, galaxies undergo a phase of compaction, which slows down and eventually halts. At the same time, the sSFR shows a mild decay along at $z=4$ to $z=1$, followed by a sudden quenching phase, which drives galaxies to the quiescent state. Both the end of the compaction and the beginning of the quenching phases coincide at $z=1.0$ to $z=0.8$ interval, during the green valley phase. Figure \ref{plot_evol_logA} shows that, despite sharing a similar beginning, the compact Illustris green valley galaxies undergo an intensified version of the same compaction and quenching processes experienced by their normal-sized counterparts.

\subsubsection{Individual evolution of simulated normal-sized and compact galaxies}\label{individual_evolution}

Median values are suitable to show general trends, although they hardly inform about the mechanisms responsible for the star formation quenching. Tracking individual galaxy histories allows us to connect simultaneous parameter fluctuations to shed light on the underlying quenching mechanisms. Individual galaxy histories are diverse, due to the random nature of their interactions, and they should not be considered more representative individually. Our goal with the simulations is to understand why more compact galaxies tend to quench faster than their normal-sized counterparts, as observed in Figure \ref{sfh_illustris_median_evolution}.

We select two examples: a normal-sized and a compact green valley Illustris galaxies, which best match the median values shown in Figure \ref{plot_evol_logA}. The Illustris sub-halo identifications (at $z = 0.8$) for these two examples are $\text{id} = 67765$ and $\text{id} = 248901$, for compact and normal-sized galaxies, respectively. Figure \ref{evolution_of_parameters_individual_evolution} shows the parameter evolution for the compact (blue) and normal-sized (red) galaxy examples along the last 12.2 Gyrs (from $z=4$ to $z=0$), from progenitors to descendants along their merger-trees. The shaded vertical bands mark the merger events with major impact on the resembling coloured galaxy. Selected parameters are: (a) the SFR and central super massive black hole (SMBH) accretion rate; (b) the gas and stellar masses; (c) the {\it in-situ} and {\it ex-situ} stellar masses. In the Illustris terminology, in-situ stars are those formed in the current galaxy from accreted gas, while ex-situ stars are accreted after being formed in other galaxies \citep{Rodriguez-Gomez2016}. Panel (d) represents the compactness ($A$), as defined in Equation \ref {compactness_definition_equation}. Note that higher $A$ values correspond to more diffuse structures. Panel (e) shows $\Sigma_{1kpc}$ the stellar mass density in the central 1 kpc while panel (f) represents the central SMBH mass evolution. Finally, panel (g) shows the evolution of the average gas temperature in a circumgalactic spherical shell comprised between 40 and 60 kpc. The typical temperature profiles, not shown, follow a bubble-like pattern, with low gas temperatures at the central 10 kpc sphere, followed by a broad temperature rise outside 10-20 kpc. The selected shell used for this plot represents the high temperature bubble. Below we aim to detail plausible quenching scenarios based on the evolution of the aforementioned parameters in Figure \ref{evolution_of_parameters_individual_evolution}.  

According to Figure \ref{plot_evol_logA} both compact and normal-sized galaxies exhibit two main phases: the {\it compaction} phase, in which the system becomes increasingly compact (towards lower $A$ values) while keeping a high and slightly decreasing sSFR; and the {\it quenching} phase, right after the compactness has reached its maximum at $z \sim 1$, in which SF is rapidly quenched. By tracking compact galaxy evolution (panel b in Figure \ref{evolution_of_parameters_individual_evolution}), we see that along the compaction phase, from $z=4$ to $z=1.0$, both stellar and gas masses steadily increase, as a result of minor mergers. The merger history of each galaxy is accessible from the simulation's merger tree (not displayed here) and shows that earlier accretion is characterised by gas-rich mergers, with gas-to-star mass ratios larger than 3. Panel (c) shows that earlier than $z=1$ the mass of the in-situ stellar component is typically three times larger than that of the ex-situ component, supporting the finding that mergers are very gas rich, contributing few ex-situ stars. Once accreted into the galaxy, gas is efficiently converted into in-situ stars, as made evident by the high SFR (panel a in Figure \ref{evolution_of_parameters_individual_evolution}). Three important mergers take place at around $z=3.3, 1.3 \text{ and } 0.3$ (bluish shaded bands). The first one, with a large gas-to-star mass ratio of 5, has a long-lasting impact on the galaxy structure, triggering a rapid increase in the central compacity (panel e), the mass of the SMBH (panel f) and the SFR (panel a). Oppositely, the third merger at $z=0.3$ with a galaxy of comparable mass has much lower structural impact due to a low gas-to-star mass ratio of 0.3. The second merger, at $z=1.3$, is the most relevant for the present study because it prompts the end of the compaction phase and the beginning of the SF quenching, 1 Gyr later at $z=1$. This is a relatively gas-rich merger with gas-to-star mass ratio of 0.9, which means the addition of a significant gas mass ($3.0 \times 10^{10}$ M$_{\odot}$). Initially, the merger triggers an intense SF burst, with SFRs higher than 100 M$_{\odot}$ yr$^{-1}$ (panel a), accompanied by a further structural compaction increase (panel d). The sharp peak in the SMBH accretion (panel a) shows that a fraction of the gas is funneled to the galaxy centre, increasing both the SMBH mass (panel f) and the central density (panel e). Initially, the gas accretion towards the galaxy centre induces both a starburst and the AGN ignition. The starburst increases both the core density and galaxy compactness until eventually the AGN kinematic-mode outflows deplete the surrounding gas, accelerating the SF quenching \citep[e.g.,][]{Kocevski2017}. Along the quenching phase, starting at $z\sim1$, only the late gas-poor merger at $z=0.3$ perturbs the progressive path to quiescence. The prominent SMBH accretion peak at $z\sim0.6$ (panel a) is probably spurious, as it is neither connected to any merger nor accompanied by any SMBH mass variation. 

The normal-sized galaxy ($\text{id} = 248901$) also exhibits compaction and quenching phases. Except for the higher $A$ value, this galaxy should be rather similar to the previous compact example, because they are selected to be at the green valley by $z=0.8$, with M$_{\star}>10^{10.7}$~M$_{\odot}$. However, as shown in Figure \ref{evolution_of_parameters_individual_evolution}, their progenitors at $z=4$ differ in almost any studied aspect, with the normal-sized galaxy, e.g., much more diffuse ($A\sim600 \text{ vs. } A\sim40$) and with two orders of magnitude less mass in the central part than that in compact one. The key to its catching-up with the compact galaxy at $z=0$ is a prominent gas-to-star mass ratio (panel b) combined with a consistently high SFR (panel a), which results in an efficient in-situ star formation (panel c). In contrast, with the structural recovery of the normal-sized galaxy, its central part evolution is kept rather ineffective, as shown by the three central parameters, the SMBH accretion (panel a), the SMBH mass (panel f) and the density in the central 1 kpc (panel e). Except for the early gas-rich merger at $z=2.7$, the smaller mergers at $z\sim1.4 \text{ and } z\sim0.2$ had a minor impact on the galaxy. Nevertheless, the gas-rich merger at $z\sim1.4$ funnels a small fraction of gas to the galaxy centre (panels a, e and f) inducing both a starburst, evidenced by a slow central density increase and the AGN ignition, which triggers the quenching phase. 

The Illustris simulation discriminates between the kinematic and thermal AGN-feedback modes \citep[e.g., ][]{Genel2014, Haider2016}. In the thermal-mode, also called quasar-mode, a fraction of the energy released by the AGN is used to thermally heat the gas cells around the black hole. Alternatively, in the kinematic-mode, also called mechanical- or radio-mode, the AGN launches jets that inject energy into bubbles of the circumgalactic and halo gas, preventing gas cooling and originating massive gas flows. The thermal-mode is more effective at larger BH accretion rates in galaxies with stellar masses below $\sim10^{10.5}$~M$_{\odot}$, being the predominant AGN feedback mode along the intense accretion phase, at redshifts higher than $\sim$ 2. On the other hand, the kinematic-mode predominates in massive galaxies at lower redshifts, in a low accretion regime.

We have worked out the evolution of the radial gas temperature profile of the two galaxies. As stated earlier, typical temperature profiles, not shown, follow a bubble like pattern, with low gas temperatures at the central 10 kpc sphere, followed by a broad temperature rise after 10-20 kpc. These profiles suggest the action of a kinematic-mode for our two prototype galaxies. Nevertheless, central temperatures bursts are also observed, generally accompanying circumgalactic heating, suggesting coexisting thermal-mode feedback episodes. According to several studies \citep[e.g., ][]{Barai2014, Choi2015}, the kinematic-mode AGN feedback is much more efficient at quenching star formation than the thermal-mode, so we have worked out the evolution from z=4 to z=0 of the average circumgalactic gas temperature in a 40-60 kpc spherical shell, where the kinematic-mode is supposed to deposit its energy. This is shown in Figure \ref{evolution_of_parameters_individual_evolution} (panel g).

Gas temperature variations for both galaxies show only a weak correlation with the SMBH accretion (panel a), as expected by the low BH accretion threshold needed to activate the kinematic-mode. It is worth noticing that along the critical redshift interval (z$\sim$1.0-0.8), in which SF quenching takes place, the circumgalactic temperature drops for the compact galaxy, while it rises around the normal one (Figure \ref{evolution_of_parameters_individual_evolution} (panel g)). Additionally, we have measured a gas loss of M$_{gas} = 1.4\times 10^{10}$~M$_{\odot}$ in the central 5 kpc sphere of the compact galaxy along the z=1.0-0.8 redshift interval, compared to a M$_{gas} = 1.1\times 10^{10}$~M$_{\odot}$ and M$_{stellar} = 3.6\times 10^{9}$~M$_{\odot}$ increase in its normal-sized counterpart. These evidences point to gas depletion, via massive gas outflows, as the source of the star formation quenching in the compact galaxy. On the other hand, the normal galaxy shows a rising circumgalactic gas temperature together with a central stellar density and total gas mass increase, a sign of a slow quenching via {\it strangulation}, in which the star formation persists inside a hot bubble until the cool gas reservoir has been consumed.

In  summary, the key difference between the compact and normal-sized galaxy SF quenching lies in the AGN feedback efficiency. First of all, compact galaxies appear to be particularly effective at igniting the AGN by driving cold gas to the galaxy centre. In addition, the more massive compact galaxy BH prompts an AGN kinematic-mode feedback, which depletes the gas and efficiently quenches the star formation. Besides the stellar feedback from stellar winds and supernovae explosions, this is probably the key to understanding the different quenching rates found in Figure \ref{sfh_illustris_median_evolution}.

\section{Discussion and Conclusions}\label{conclusions_and_discussion}

\subsection{Formation and evolution of compact galaxies}

Our analysis using Illustris simulations indicates that compact galaxies may show strong stellar and AGN feedback during the blue nugget phase, which is in agreement with previous works \citep[e.g., ][]{Dekel2014, Zolotov2015}. Next, we describe the possible scenario for the formation and evolution of compact galaxies at high redshifts.

\subsubsection{Blue nugget phase}

High gas fraction typically found in high-$z$ galaxies can cause a violent disk instability \citep[VDI, ][]{Dekel2014}, i.e., strong gravitational instabilities in the galactic disk, which results in the formation of giant clumps that rapidly migrate towards the centre of the host galaxy. A compact star-forming galaxy is formed, with a very high star formation rate \citep[$\sim100$ M$_{\odot}$ yr$^{-1}$, ][]{Zolotov2015}. We find evidences in our simulated galaxies that gas-rich minor mergers can also accelerate the compaction process. This is in agreement with \citet{Zolotov2015}, which argue that, besides VDI and minor mergers, counter-rotating streams \citep[streams from the cosmic web which are counter-rotating in relation to the existing disk, ][]{Danovich2015}, low angular momentum recycled gas \citep[gas which was earlier ejected by the galaxy and returned to the disk, ][]{Elmegreen2014} and tidal compression \citep[tidal forces within satellites galaxies that push gas towards the satellite galaxy centre, ][]{Dekel2003} can also drive the blue nugget formation.

\subsubsection{Green nugget phase}

In this phase the galaxy's star formation activity is reduced (quenching phase) and the global colours place the galaxy within the green valley in the colour-magnitude diagram. The green nugget phase may be triggered by one or a combination of several mechanisms including stellar feedback \citep{Murray2005,Lagos2013}, quasar-mode feedback from an AGN \citep{DiMatteo2005, Ciotti2007, Schawinski2009, Ishibashi2017} and morphological quenching \citep{Martig2009}. These mechanisms are associated with short time-scales and, hence, with a \textit{fast mode} of quenching. Recently \citet{Kocevski2017} found that $\sim40\%$ of blue nuggets host an AGN at $z>1$, providing further support to this picture.

\subsubsection{Red nugget phase}

Finally, a compact quiescent galaxy is formed. Additional mechanisms can act in \textit{maintenance mode} (preventing new star formation activity), such as virial shock heating and radio mode AGN feedback \citep{Croton2006, Dekel2006, Cattaneo2009}.

\subsection{Comparison with previous works}

The formation of compact galaxies, and consequently the evolution of their internal properties are more likely to occur at high redshift, when larger reservoirs of cold gas are available for star formation \citep{Tacconi2010, Genzel2010}. \citet{Barro2013} discuss that, in general, galaxies evolve via one of two principal tracks: the \textit{early-track}, that happens at early times ($z=2-3$) and; the \textit{late-track}, that is more common at late times ($z=0-2$). In the early-track the main mechanisms involved are the ones described by the \citet{Dekel2014} model, whereas in the late-track star formation is quenched at a slower pace in green valley galaxies with normal-sized extensions. However, the rapid quenching time-scales that we measure for our compact and ultra-compact green valley galaxies suggest that although early-track red sequence formation was likely more significant at higher redshifts, it can still act at $z<1$. Moreover, Figure \ref{coadded_galaxy_spectra_and_h_delta_vs_dn4000_for_compactness_study} shows that the spectral index H$\delta_A$ (associated with a burst of star formation in the recent past) is much stronger in compact and ultra-compact green valley galaxies than in those in normal-sized ones, indicating that green nuggets are moving more abruptly from a star-forming to a quiescent state, when compared to the general population. The H$\delta_A$ values we measure for our green nuggets are in agreement with those found in \citet{Barro2016} at $z\sim1.7$, suggesting that the same processes that are forming compact galaxies at $z>1.0$ are likely acting as well at $0.5<z<1.0$. 

We exploit the Illustris simulation to compare our observational results based on COSMOS green valley galaxies. In qualitative agreement the Illustris simulation shows that green nuggets quench their star formation $\sim2$ times faster than normal-sized galaxies. We note, however, that the exponential indices ($\gamma$ [Gry$^{-1}]$) in simulated green valley galaxies (i.e.,  0.53 and 1.11 for normal-sized and compact galaxies, respectively) are shorter than those measured for real green valley galaxies (i.e.,  2.73, 6.17 and 9.65 for normal-sized, compact and ultra-compact galaxies, respectively). To better understand this discrepancy and more accurately verify if predictions by simulations are in agreement with observations, we compare the number density evolution of compact star-forming galaxies and compact quiescent galaxies in the Illustris simulation with those estimated by \citet{Barro2013} based on compact galaxies within the CANDELS fields. 

\begin{figure}
  \includegraphics[width=\columnwidth]{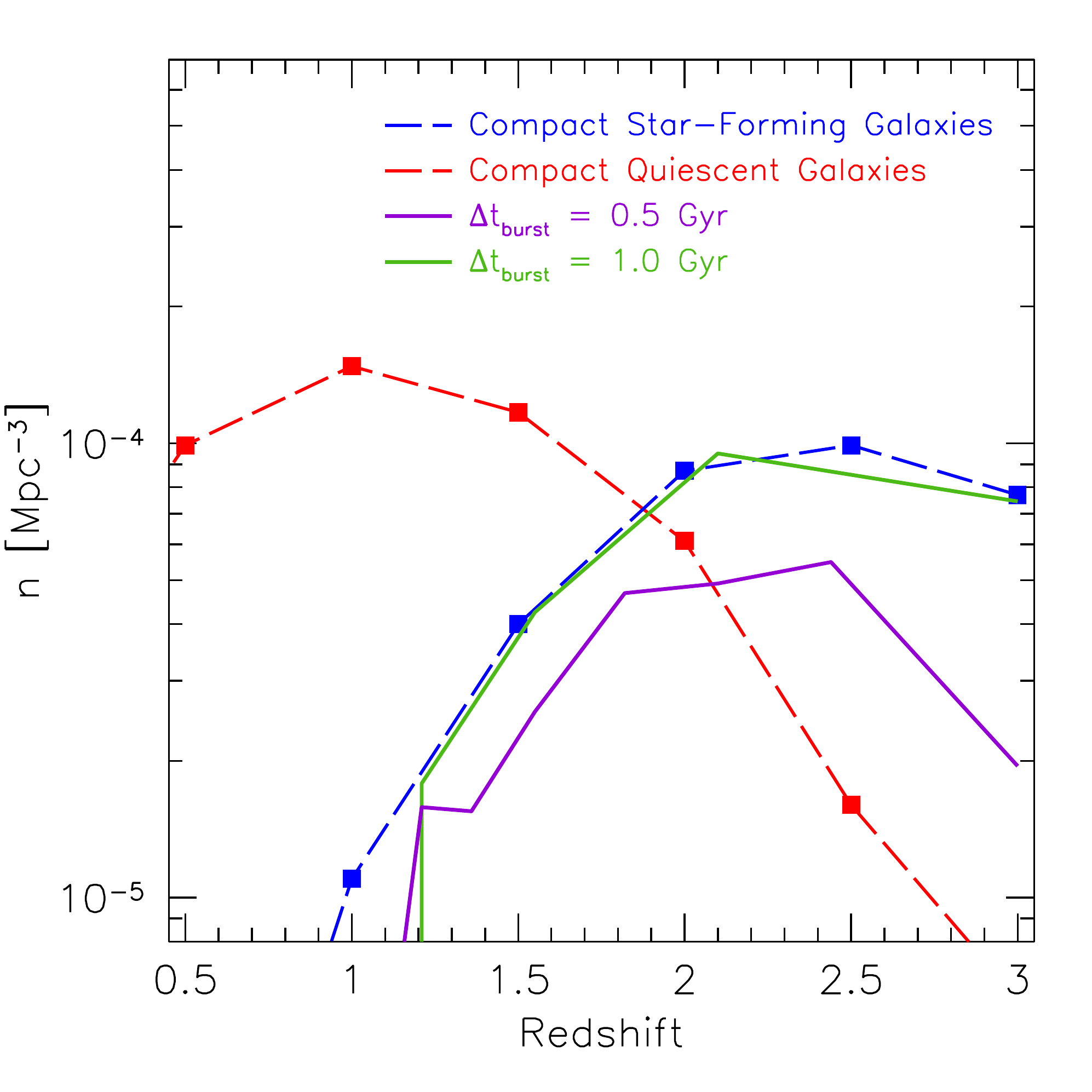}
  \caption{Number density evolution of compact star-forming galaxies, cSFGs (blue dashed line) and compact quiescent galaxies, cQGs (red dashed line) for our Illustris sample. The green and purple continuous lines correspond to models of how fast cSFGs quench into cQGs. The argument is that both number density curves are not independent but connected by the fact that a fraction ($\sim$0.6) of the cSFGs end up quenched as cQGs after a certain time interval $\Delta t_{burst}$. The equation $0.6 \times n_{SFG}(t) = n_{QG}(t+\Delta t_{burst}) - n_{QG}(t)$ allows the calculation of $\Delta t_{burst}$. The best match with the $n_{SFG}(t)$ curve is found with $\Delta t_{burst}$ = 1 Gyr (green line).}
  \label{fit_number_density_nuggets_as_a_function_of_redshift}
\end{figure}

\citet{Barro2013} use a slightly different compactness criterion, M$_{\star}$/R$_{eff}^{1.5} > 10.3$ M$_{\odot}$ kpc$^{-1.5}$, as well as a stellar mass cut at M$_{\star} > 10^{10}$ M$_{\odot}$ and a $\text{log(sSFR)}=-0.5~\text{Gyr}^{-1}$ threshold to separate star-forming from quiescent galaxies. We imposed these criteria to carry out a similar selection using Illustris galaxies and consider the number density evolution of compact star-forming galaxies and compact quiescent galaxies in Figure \ref{fit_number_density_nuggets_as_a_function_of_redshift}, in an attempt to reproduce \citet{Barro2013} study. In their Figure 5, \citet{Barro2013} represent separately their results on the number density evolution of both compact quiescent (cQGs) and compact star-forming (cSFGs) galaxies along the redshift interval $z=1.4 \text{ to } z=3.0$. According to the authors, these evolutionary paths are not independent, but connected by the fact that all the cSFGs end up quenching into cQGs after a certain time interval $\Delta t_{burst}$, measured in Gyrs. A simple evolutionary model can be used to recover the best fitting $\Delta t_{burst}$. Their simple model is summarized in the equation, $n_{SFG}(t) = n_{QG}(t+\Delta t_{burst}) - n_{QG}(t)$, stating that any $n_{QG}$ increase along a $\Delta t_{burst}$ Gyr period is originated by quenched cSFGs. They find a median value of $\Delta t_{burst}$ = 0.8 Gyr, in agreement with our result for compact green valley galaxies in Section \ref{median_evolution}. We apply the same analysis to the Illustris compact population and find that the simple model of \citet{Barro2013} does not provide a good match for the observed number density evolution for the Illustris galaxies. A match in Illustris is only possible when the requirement stating {\it all compact star-forming galaxies quench into compact quiescent galaxies} is relaxed to {\it a fraction of the compact star-forming galaxies quench into compact quiescent galaxies}. By tracking the compactness-sSFR evolution of all the individual cSFGs, we find that although all compact star-forming galaxies quench into quiescent galaxies, roughly one third of them reduce their compactness along the process and are thus excluded from the compact quiescent galaxy sample. In summary, we have modified the simple \citet{Barro2013} equation introducing $0.6 \times n_{SFG}(t)$ on the left side of the equation. After that correction, we get the best fit with $\Delta t_{burst}$ = 1 Gyr, of the order of the quenching timescale $(1/\gamma)$ = 0.8 Gyr, found for the Illustris compact green valley galaxies in Section \ref{median_evolution}.

\subsection{Quenching time-scales comparison between real and simulated green nuggets}

Despite the required modification of the original \citet{Barro2013} equation, we conclude that the Illustris predictions on the number density evolution of cSFGs and cQGs are in reasonable agreement with the observations and, therefore, the absolute quenching time-scales in simulations are adequate. The disagreement between these quenching time-scales with those found in COSMOS galaxies can be attributed to the the NUV$-r$ colours. Our method for measuring quenching time-scales (Section \ref{methodology}) requires the exponential index $\gamma$ to be a function of the spectral indices, D$_n$(4000) and H$\delta_A$, and the NUV$-r$ colour. Although the spectral indices are not affected by galaxy-intrinsic extinction, the NUV$-r$ colour is strongly affected, specially in starburst regions \citep[e.g., ][]{Calzetti1994, Calzetti2000}. As described in Section \ref{extinction_correction}, we apply the \citet{Salim2007} models just to remove dusty star-forming galaxies from the green valley. As mentioned before, the procedure to correct dust reddening works well with blue galaxies but fails in green valley and red sequence galaxies. We proceed to estimate quenching time-scales in our final green valley sample considering the observed NUV$-r$ colours (without dust extinction correction). Model green valley locations in the H$\delta_A$ $-$ D$_n$(4000) diagram (i.e., black dots on coloured curves, bottom panels in Figure \ref{coadded_galaxy_spectra_and_h_delta_vs_dn4000_for_compactness_study}) shift down when redder colours are considered. By using observed NUV$-r$ colours (without dust extinction correction) for our green valley galaxies, we are potentially using H$\delta_A$ $-$ D$_n$(4000) model grids that are systematically shifted down. The impact of this is a potential overestimate in the final exponential quenching index ($\gamma$) in real green valley galaxies. In future works, we will address this issue carefully \citep[e.g., through the use of Balmer decrement, ][]{Calzetti1994}. 

However, we warn the reader that the procedure to estimate quenching time-scales in COSMOS green valley galaxies is model dependent, as discussed in \citet{Nogueira-Cavalcante2018}. Therefore, this methodology is properly applied when we compare different quenching time-scales from different class of green valley galaxies, i.e., for this particular work, when one compares quenching time-scales between normal-sized and compact (ultra-compact) green valley galaxies.

\subsection{Evolution of green nuggets with galaxy colour and redshift}

\begin{figure*}
\includegraphics[width=\textwidth]{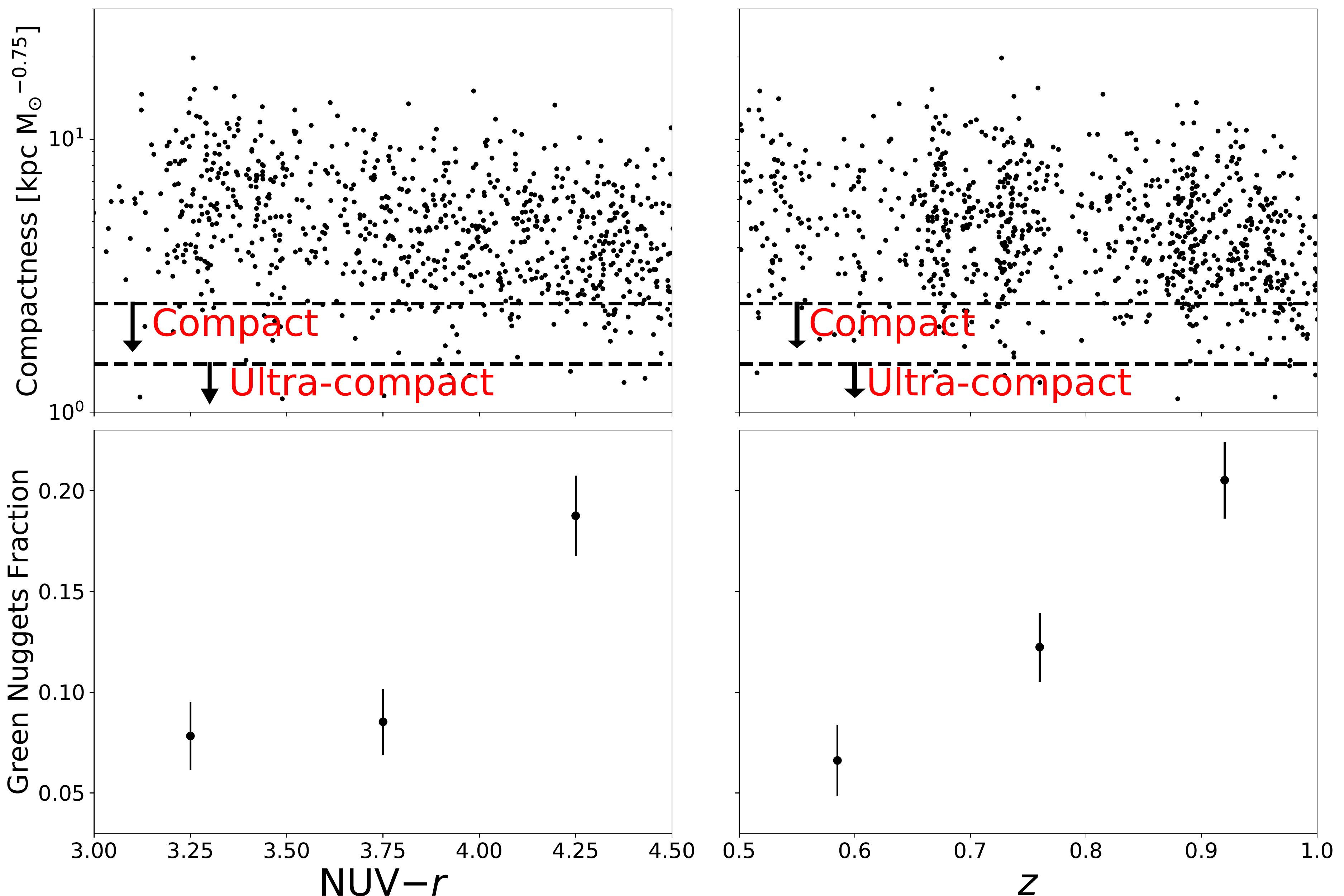}
\caption{\textbf{\textit{Top:}} Compactness as a function of NUV$-r$ colour (left panel) and redshift (right panel) of green valley galaxies in our sample. The black arrows in each panel indicate the direction for higher compaction (i.e., the smaller the compactness value, the more compact the galaxy). The top dashed line delimits compact galaxies from normal-sized ones and the bottom dashed line delimits ultra-compact galaxies from the compact ones. \textit{\textbf{Bottom:}} Green nugget fraction as a function of NUV$-r$ colour and redshift. We define bins to have approximately the same number of galaxies in each one.}
\label{compactness_as_a_function_of_color_and_redshift}
\end{figure*}

We also analyse the distribution of compactness as a function of NUV$-r$ colour and redshift in order to study the evolution of the fraction of green nuggets with cosmic time and as a function of position within the green valley (Figure \ref{compactness_as_a_function_of_color_and_redshift}). We find that the fraction of green nuggets increases with galaxy colour, indicating that the majority of compact objects occupy the redder part of the galaxy CMD at $0.5<z<1.0$. We also identify a clear trend with redshift, where the green nugget population becomes more significant at early times, going from $\sim5$\% at $z\sim0.6$ to $\sim20$\% at $z\sim0.9$. These results are consistent with the observed increase in the blue nugget fraction at higher redshifts \citep{Barro2013,Charbonnier2017} and suggest that the transformation of blue nuggets into red ones $-$ and hence the prevalence of green nuggets $-$ was more common at higher redshifts. The observed decrease in green nugget number density as we consider lower redshifts and redder colours suggests that progenitors of green nuggets (i.e., blue nuggets) are more abundant at higher redshifts and decrease in number density at later cosmic times.  

At the redshifts of our study, $0.5<z<1.0$, the majority of compact galaxies are passive $-$ that is, most nuggets become red nuggets at early epochs via the early-track red galaxy formation. We identify an increase in the fraction of green nuggets towards the far-side of our redshift range, which indicates that transitional green compact galaxies become more common at earlier times. Consistent with the early-track red sequence formation scenario, at early times we expect galaxies to be more gas-rich and, consequently, more prone to disk instabilities, compaction and rapid quenching, following the scenario proposed by \citet{Dekel2014}. However, the rapid quenching time-scales that we measure for our compact and ultra-compact green valley galaxies suggest that although early-track red sequence formation was likely more significant at higher redshifts, it can still act at $z<1$.

\section{Summary}\label{summary}

The existence of compact, massive, passive galaxies $-$ commonly known as red nuggets $-$ and their evolution still presents challenges to galaxy formation models. In an effort to study star formation quenching in galaxies that give rise to such a population we exploit deep imaging and ample spectroscopy of COSMOS field galaxies at $0.5<z<1.0$ to identify so-called green nuggets and quantify the time-scale of their transition through the green valley. We divide our sample of COSMOS green valley galaxies into ultra-compact, compact and normal-sized galaxies based on the stellar masses provided by the UltraVISTA catalogue and angular sizes of the ZEST catalogue. We find that the star formation quenching time-scales in green nuggets (ultra-compact and compact galaxies) are much shorter than in normal-sized galaxies. We complement our study by investigating the evolution of physical parameters (e.g., star formation history, AGN feedback) in simulated galaxies from the Illustris project that match our adopted compactness criteria. Our results both from real and simulated data are consistent with scenarios of violent disk instability and gas-rich mergers that can cause a rapid transition from a compact star-forming galaxy (blue nugget), followed by a quenching phase (green nugget), very likely caused by strong AGN feedback, which depletes central gas and/or injects energy in a circumgalactic hot gas bubble, strangulating the SF process. Although at the redshifts of our study ($0.5<z<1.0$) the majority of compact galaxies are already passive $-$ that is, most nuggets are red nuggets formed via early-track at early epochs the formation of red nuggets, the rapid quenching time-scales that we measure for our compact and ultra-compact green valley galaxies suggest that although early-track red sequence formation was likely more significant at higher redshifts, it can still act at $z<1$. 

\section*{Acknowledgements}

We thank the anonymous referee for the very helpful report, which helped us clarify several important points from the original manuscript. JPNC was supported by a PhD grant from CAPES and an Institutional Capacity Program grant from CNPq. TSG and KMD thank the support of the Productivity in Research Grant of the CNPq. IGdlR acknowledges a grant from the Spanish Ministry of Education, Culture and Sports, in the frame of its program to promote the mobility of Spanish researchers to foreign centers. IGdlR also acknowledges support from grant AYA2016-77237-C3-1-P from the Spanish Ministry  of  Economy  and  Competitiveness (MINECO). AC acknowledges support by the Brazilian Science Without Borders program, managed by CAPES and CNPq. 


\bibliographystyle{mnras}
\bibliography{References_by_Nogueira-Cavalcante}


\bsp	
\label{lastpage}
\end{document}